\documentclass[11pt]{iopart}
\usepackage{bm}
\usepackage{hyperref}
\usepackage{amssymb}
\usepackage{feynmp}
\usepackage{amsopn}
\usepackage{setstack}
\usepackage{graphicx}
\usepackage{slashed}
\RequirePackage{ifpdf}

\ifpdf
\else 
\fi

\newcommand{\Mew}{M_{\mathrm{EW}}}

\newcommand{\Msusy}{M_{\mathrm{SUSY}}}

\newcommand{\thetaw}{\theta_{\mathrm{W}}}
\newcommand{\sw}{s_{\mathrm{W}}}
\newcommand{\cw}{c_{\mathrm{W}}}
\newcommand{\tildesw}{\tilde{s}_{\mathrm{W}}}

\newcommand{\figlabel}[1]{\textsf{({#1})}}
\newcommand{\wavefn}{\mathsf{Z}}

\renewcommand{\d}{\mathrm{d}}
\newcommand{\vect}[1]{\bm{\mathrm{{#1}}}}

\newcommand{\im}{\mathrm{i}}

\DeclareMathOperator{\Ei}{Ei}

\newcommand{\subtitle}[1]{\textsf{\textbf{{#1}}}}

\makeatletter
\newcommand\numberwithin[2]{\@addtoreset{#1}{#2}}
\makeatother
\numberwithin{footnote}{section}

\begin{document}
\begin{fmffile}{diags}
        \begin{flushright}
                \textsf{DESY 08-207}
        \end{flushright}
        \title{Collider constraints on interactions of dark energy with the
        Standard Model}
        
        \author{Philippe Brax$^{1}$, Clare Burrage$^2$, Anne-Christine Davis$^3$,
        \\ David Seery$^3$ and Amanda Weltman$^{3,4}$}
        \vspace{3mm}
        
        \address{$^1$ Institut de Physique Th\'{e}orique,
          CEA, IPhT, CNRS, URA2306, F-91191 Gif-sur-Yvette c\'{e}dex,
          France \\[3mm]
        $^2$ Theory Group,  Deutsches Elektronen-Synchrotron DESY,  D-22603,
          Hamburg, Germany \\[3mm]
        $^3$ Department of Applied Mathematics and Theoretical
          Physics \\
          Centre for Mathematical Sciences, University of Cambridge, \\
          Wilberforce Road, Cambridge, CB3 0WA, United Kingdom \\[3mm]
        $^4$ Department of Mathematics and Applied Mathematics, 
          University of Cape Town, Private Bag, Rondebosch, South Africa, 7700}
        \vspace{3mm}

        \eads{\mailto{djs61@cam.ac.uk}}
        
        \begin{abstract}
                We study models in which
                a light scalar dark energy particle couples to the
                gauge fields of the electroweak force, the
                photon, $Z$, and $W^\pm$ bosons. Our analysis applies to a
        large class of interacting dark
                energy models, including those in which the dark energy
                mass can be adjusted to evade fifth-force bounds by the
                so-called ``chameleon'' mechanism.
                We conclude that---with the usual choice of Higgs sector---%
                electroweak precision observables are screened
                from the indirect effects of dark energy,
                making such corrections effectively
                unobservable at present-day colliders, and limiting
                the dark energy discovery potential of any future International
        Linear Collider.
                We show that a similar screening effect applies to processes
                mediated by flavour-changing neutral currents, which can be
                traced to the Glashow--Iliopoulos--Maiani mechanism.
                However, Higgs boson production at the Large Hadron Collider
                via weak boson fusion may receive observable corrections.
        \vspace{3mm}

        \begin{flushleft}
        \end{flushleft}
        \end{abstract}
        \maketitle

        \section{Introduction}
        \label{sec:introduction}
        
        The emergence of cosmology as a data-driven science in the late 1990s
        enabled our theories of the universe to be promoted from mostly
        speculation to meaningful quantitative investigation.
        Although many components of what now forms the standard
        ``concordance'' $\Lambda$CDM cosmology had been proposed
        prior to the quantitative revolution
        and were found to be consistent with
        experiment, among the more surprising revelations
        was the emergence of a new scale at
        around $10^{-3}$~eV, associated with an apparent acceleration
        of the cosmological expansion.
        The properties of Nature at this scale have been accessible since
        the earliest days of particle physics, and our models of
        microscopic processes at these energies are now very well tested.
        It was therefore surprising to discover that this hitherto
        mundane scale was to be associated with an exotic species of matter     
        with
        energy density $\Lambda \sim (10^{-3} \, \mbox{eV})^4$ and equation
        of state $p \approx - \rho$.
        
        The microphysics associated with this energy density
        remains unknown.
        The most parsimonious interpretation of the data requires only
        Einstein's ``cosmological term,'' which we now know to be
        degenerate with the
        aggregate effect of quantum vacuum fluctations.
        At a mass scale $M$ these contribute a cosmological
        effect of order $M^4$.
        Consequently, if we take our well-tested quantum-mechanical theories of
        physics seriously even at comparatively modest scales
        (up to $\Mew \sim 100 \, \mbox{GeV -- 1 TeV}$) then
        without a remarkable cancellation we encounter
        a serious disagreement with the data.
        An alternative interpretation is to suppose that unknown
        physics renders the quantum zero-point energy negligible or
        unobservable. If this is the case, it is possible
        that our present phase of acceleration
        is driven by the potential
        energy associated with some scalar field. This
        field would have to be very light on large scales in our present
        vacuum, with mass of order $H_0 \sim 10^{-33}$ eV,
        but it might evade the stringent
        bounds associated with long-range forces mediated by light bosons
        if its mass could be adjusted to be large in regions of high
        average density. Theories of this type were proposed by Khoury \&
        Weltman \cite{Khoury:2003aq,Khoury:2003rn}, who called such fields
        ``chameleonic'' in view of their ability to vary their properties
        depending on the environment.%
                \footnote{For earlier work, see
                Refs.~\cite{Mota:2003tm,Clifton:2004st}.}
        
        The chameleon property means that models involving these fields can
        give rise to successful acceleration at late times
        \cite{Brax:2004qh}, while remaining
        consistent with known constraints on long-range physics. Such
        models are attractive for another reason, because the requirement
        that the field can respond to local variations in the density of
        bulk matter means that couplings to Standard Model states are
        mandatory. Chameleonic fields are therefore constrained by precision
		measurements of the early universe---in particular,
		observations of Big Bang Nucleosynthesis (`BBN') and the redshift of
		recombination \cite{Brax:2004qh,Mota:2006fz}.
		As the universe cools the background
		dark energy field remains fixed in the minimum of its potential,
		whose location slowly drifts. The result is
		a variation in
		the mass of any particle to which dark energy is coupled.
		However, acceptable models are constructed in such a way that
		only small changes in particle mass can be expected,
		and therefore the constraints from observations such as
		BBN are rather weak.
		Interesting bounds have also been obtained
		from a variety of astrophysical
        and terrestrial processes
        \cite{Brax:2007ak,Brax:2007hi,Brax:2007vm,Burrage:2007ew,Brax:2008hh,
        Chou:2008gr,Gies:2007su,Ahlers:2007st,Burrage:2008ii}.

        These couplings also imply
        the existence of an interesting collider phenomenology.
		With the aim of complementing the cosmological and astrophysical
		tests,
		our purpose in this paper is to
        take the first steps towards understanding the
        implications of dark energy corrections for Standard Model processes
        which can be observed at present and future particle colliders.
        A related study has been performed by Kleban \& Rabadan
        \cite{Kleban:2005rj}.
        
        What form would these corrections take?
        We expect that the dark energy scalar is not charged under any of the
        usual gauge quantum numbers associated with the Standard Model.
        Its couplings to Standard Model states are therefore
        unrestricted by considerations of gauge invariance.
        Nevertheless,
        because bulk mass in the macroscopic world is
        dominated by hadrons
        it seems unavoidable for a chameleonic scalar to
        couple to those degrees of freedom charged under QCD, namely
        the quarks and gluons. Unfortunately, hadron interactions in QCD
        are non-perturbative in nature and are difficult to study.
        It is less obvious that the dark energy is obliged to couple to degrees
        of freedom charged under the electroweak $SU(2) \times U(1)$
        gauge symmetry, but if it does then one might imagine that
        such interactions would offer a more tractable
        probe of the theory than the complicated colour physics of QCD.
        Our purpose in this work is to study the comparatively clean
        experimental signatures which arise at low energy from the
        existence of couplings between dark energy and Standard Model states
        which carry electroweak quantum numbers.

        Interactions between a scalar dark energy species and the
        electroweak sector need not be harmless.
        For example, variation in
        the dark energy vacuum expectation value could lead to a shifting
        fine-structure constant or loss of conservation of electric charge
        \cite{Carroll:1998zi,Shaw:2005ip}.
        From the perspective of collider phenomenology, there is another
        serious difficulty: fundamental scalar fields
        are well-known to depend sensitively on the details of physics
        in the ultra-violet. If Standard Model particles can radiate into
        light chameleon states while participating in some measurable process,
        then we must allow for the possibility of
        significant corrections to observable Standard Model reactions.
        Indeed, it is a serious question whether
        \emph{any} dark energy model of this
        type can be compatible with existing data.
        It is also important to understand whether we should expect
        dramatic signals at impending high-precision experiments such as the
        \emph{Large Hadron Collider} (LHC) at
        the European Organization for Nuclear Research (CERN) or at
    	a proposed future International Linear Collider. 

        In this paper, we study the effect of such radiative corrections.
        Our results apply to models of chameleon dark energy,
        and also to alternatives such as coupled quintessence, or
        any beyond-the-Standard-Model scalar species which is light
        in the laboratory environment. Similar issues have been addressed
        previously by Einhorn \& Wudka \cite{Einhorn:1992dx}, who
        determined the criteria for \emph{heavy} scalar particles to be
        screened. However, our results are not contained in their
        analysis because the scalar particles which can cause successful
        cosmological acceleration must ordinarily be very light
        in the laboratory environment
        compared to the electroweak scale, with masses of order
        $\lesssim 10^{-8}$ eV or lighter.
        It is the effect of highly suppressed couplings in the laboratory
		which allows such particles to have evaded detection, rather than the
        significant energy cost of producing them in collisions.

        In \S\ref{sec:uv} we give a brief summary of ultra-violet effects
        in scalar field theories, before going on to review the formalism
        used to study corrections to electroweak precision observables
        (\S\ref{sec:ewpo}).  In \S\ref{sec:Z-decay} we study
		corrections to the width of the Z boson (a tree-level effect),
		and show that it leads to a weak constraint.
		In \S\ref{sec:oblique} we identify a class of
        loop effects which lead to stronger constraints,
		the so-called ``oblique'' corrections.
		The key quantities we require to compute them
        are the vacuum polarizations of the $W^\pm$, $Z$
        and $\gamma$ bosons, which are obtained in \S\ref{sec:polarization}.
        In \S\ref{sec:eft} we interpret these vacuum polarizations in terms
        of an effective Lagrangian which makes their physical content
        transparent.
        In \S\ref{sec:discuss} we discuss our findings and indicate how our
        results could be extended to a larger zoology of processes,
        including so-called flavour-changing neutral currents.
        In particular, in \S\ref{sec:constraints}
        we discuss the conditions under which the largest corrections
        are ``screened,'' meaning that they do not enter in any measurable
        relationship between observables. It is only when screening occurs
        that the model is automatically compatible with the simplest
        predictions of the Standard Model.
        In \S\ref{sec:future}
        we determine the constraints which can be obtained from data
        obtained by present-day colliders, and discuss the role of future
        hadron--hadron or $e^+ e^-$ colliders.  
        Finally, in \S\ref{sec:conclusions} we state our conclusions.
        Some technical details are collected in two Appendices.

        We choose units throughout such that $\hbar = c = 1$. Our metric
        convention is $(-,+,+,+)$, so that on-shell particles have negative
        invariant momenta. Spacetime indices are denoted by lower-case
        Latin indices $\{ a, b, c, \ldots \}$, and we label the species of
        vector bosons by upper-case indices $\{ A, B, C, \ldots \}$.

        \section{Electroweakly interacting dark energy}
        \label{sec:interacting-de}
        
        \subsection{Ultra-violet effects}
        \label{sec:uv}

        The problem of sensitivity to ultra-violet effects
        is universal in any theory of scalar fields.
        While it is an obstacle for model-building, UV sensitivity
        can be exploited as a tool to probe the theory at energies
        much higher than those which can physically be realized
        in particle accelerators.
        An important example of this
        occurs in the Higgs sector of the Standard
        Model, which has many parallels with the case of interacting
        dark energy. For this reason we digress to give a brief
        discussion of the Higgs case, before returning to dark energy
        in \S\ref{sec:de-interactions}.

        All particles which gain their mass via the Higgs mechanism
        are entitled to radiate into Higgs states,
        and in consequence it was pointed out long ago by Veltman that
        electroweak quantities can receive large Higgs contributions, up
        to some scale
        above which radiation is suppressed. This scale is
        presumably determined by a more complete theory of microscopic
        interactions, of which the Standard Model is an effective low
        energy limit. The Standard Model including a Higgs sector is
        precisely renormalizable, but if the Higgs is decoupled from the theory
        by taking its mass to infinity, $M_H \rightarrow \infty$,
        we should recover the divergences of the Higgsless case.
        One can therefore
        think of $M_H$ as a soft effective cutoff corresponding to the
        scale of new physics \cite{Veltman:1994vm}.
        Any large Higgs contributions must appear experimentally as deviations
        from
        the tree-level expectation, which can be summarized in terms
        of Veltman's ``$\rho$-parameter.'' In principle, this could
        receive corrections from the Higgs sector of the form
        \begin{equation}
                \rho \equiv \frac{M_W^2}{M_Z^2 \cos^2 \theta} = 1 +
                a_0 g^2 \frac{M_H^2}{M_Z^2} +
                a_1 g^2 \ln \frac{M_H^2}{M_Z^2} + \cdots ,
                \label{eq:veltman-rho}
        \end{equation}
        where $a_0$ and $a_1$ are pure numbers which must be calculated,
        $g$ is a coupling constant, and
        `$\cdots$' denotes the effect of higher-order
        radiative corrections which we have neglected.
        The current experimental constraint is
        $\rho = 1.0004^{+0.0008}_{-0.0004}$
        \cite{Amsler:2008zz}, so
        if $a_0 \neq 0$ one would obtain extremely stringent constraints
        on $M_H$. Unfortunately, in the Standard Model it turns out that
        $a_0 = 0$ \cite{Veltman:1976rt,Veltman:1977kh}, leading to a
        considerably weaker bound
        $M_H \lesssim 215 \, \mbox{GeV}$. This effect
        occurs in all Standard Model observables and has become known as the
        \emph{screening theorem}, because it protects low-energy
        observations from the effect of coupling to a large phase space
        of scalar Higgs
        states. It has been shown that the screening
        phenomenon extends to all orders in the loop expansion
        in the limit $M_H \rightarrow \infty$
        \cite{vanderBij:1983bw,Einhorn:1988tc,Einhorn:1992dx}.
        
        The same principles apply to any light scalar field.
        What happens if Standard Model particles are permitted to radiate
        into dark energy states? In the laboratory environment
        where the $W^\pm$ and $Z$ masses can be measured, the dark energy
        quanta are typically light. In this case, we must expect
        contributions to electroweak observables of the form
        described by Eq.~\eref{eq:veltman-rho}, with the Higgs mass
        $M_H$ replaced by whatever scale $M$ determines the size of the phase
        space of available states, and the coupling $g^2$ replaced
        by whatever quantity sets the interaction strength of dark energy
        with ordinary matter, which is typically a number of order $M_Z^2/M^2$.
        It then becomes extremely significant whether dark energy
        exhibits a similar screening effect, for if $a_0 \neq 0$
        then $\rho$ will generically receive corrections of $\Or(1)$.
        Such large corrections could easily lead to an unacceptable
		conflict with
        precision electroweak data. On the other hand, if the dark energy
        does exhibit screening then the corrections to $\rho$
        are roughly of order $\Or[ (M_Z/M)^{2} \ln M^2/M_Z^2 ]$
        and are therefore very small
        for any phenomenologically reasonable choice of $M$.

        We would like to emphasize that there is no reason of principle
        for the Higgs or any other scalar species to exhibit this sort of
        radiative screening. In the Higgs sector, a so-called
        ``custodial'' global $SU(2)$ symmetry becomes exact in the limit where
        the hypercharge gauge coupling $g_1$ vanishes
        \cite{Sikivie:1980hm}, which guarantees
        equality of the vector boson masses, but does not guarantee
        screening \cite{Veltman:1994vm}.%
                \footnote{In their proof that the Higgs exhibits radiative
                screening to all orders in the loop expansion,
                Einhorn \& Wudka made essential use of the $SU(2)$ custodial
                symmetry \cite{Einhorn:1988tc}. However, although the
                existence of this symmetry is \emph{necessary}, it is
                not sufficient. An integral part of of Einhorn \& Wudka's
                argument consists of a power-counting procedure entirely
                unconnected with the custodial symmetry, which determines where
                the leading divergences can appear as
				$M_H \rightarrow \infty$.}
        In the absence of any specific reason to think otherwise,
        one must imagine that a generic scalar field theory interacting
        with the electroweak sector would contribute to
        Eq.~\eref{eq:veltman-rho} with $a_0 \neq 0$.
        Although it may be possible to fine-tune a model
        of this type to be consistent with
        precision electroweak observations, this solution would be highly
        unattractive. Indeed, one would have traded an unappealing
        fine-tuning in the cosmological constant for a fine-tuning
        in the scalar model intended to replace it,
        and little would have been gained. 
        
        \subsection{The interaction Lagrangian}
        \label{sec:de-interactions}

        We will choose to work with a theory of the broken phase of the
        electroweak force in which the photon and the massive
        vector bosons $W^\pm$ and $Z$ interact
        with a single dark energy scalar $\chi$ according to the action
        \begin{eqnarray}
                \fl\nonumber
                S = -\frac{1}{4} \int \d^4 x \;
                        \Bigg\{
                                2 B(\beta \chi) (\partial^a W^{+b} - \partial^b W^{+a})
                                (\partial_a W^-_b - \partial_b W^-_a)
                                + 4 m_W^2 B_H(\beta_H \chi) W^{+a} W^-_a \\
                                \nonumber \quad \mbox{} +
                                B(\beta \chi) (\partial^a Z^b - \partial^b Z^a)
                                (\partial_a Z_b - \partial_b Z_a)
                                + 2 m_Z^2 B_H(\beta_H \chi) Z^a Z_a \\
                                \quad \mbox{} +
                                B(\beta \chi) (\partial^a A^b - \partial^b A^a)
                                (\partial_a A_b - \partial_b A_a)
                        \Bigg\} ,
                \label{eq:action}
        \end{eqnarray}
        where $W^\pm_a$ and $Z_a$ are the gauge fields associated with the
        $W^\pm$ and $Z$, respectively, and $A_a$ is the gauge field associated
        with the photon.
		Eq.~\eref{eq:action} should be thought of as an effective
		Lagrangian valid after integrating out the Goldstone modes
		of the Higgs, as emphasized by Burgess \& London
		\cite{Burgess:1992va,Burgess:1992gx}
		following earlier work
		in Refs.~\cite{Cornwall:1974km,Chanowitz:1987vj}.
		Only invariance under the electromagnetic $U(1)$ gauge group is
		required. 
        
        The quantities $m_W$ and $m_Z$ are the Lagrangian parameters
        corresponding to the mass of the $W^{\pm}$ and $Z$, which are related
        via a renormalization prescription to the physical masses $M_W$ and
        $M_Z$. In addition, we have introduced two arbitrary functions
        $B(\beta \chi)$ and $B_H(\beta_H \chi)$
        which describe how the scalar $\chi$ couples to the gauge boson
        kinetic and mass terms. These couplings are associated with
        mass scales $M \equiv \beta^{-1}$ and $M_H \equiv \beta_H^{-1}$
        (not necessarily identical%
                \footnote{Note that $M_H$ is \emph{not} the Higgs mass,
                which was discussed in \S\ref{sec:uv}
                but does not appear
                in the remainder of this paper.}%
        ) which
        control the relative strength of the interaction between
        dark energy and the weak gauge bosons, and
        between dark energy and the Higgs field respectively.
        
        Throughout this paper, we assume that the
        dark energy quanta $\chi$ have some fixed
        mass $M_\chi$, which is not subject to renormalization. This is
        tantamount to treating the entire scalar sector as an effective
        field theory, in which quantum effects have already been included,
        and for which we only wish to assess the influence of radiative
        corrections on the bare electroweak sector.
        This is
        appropriate for a phenomenological model such as a chameleon,
        which need not be a fundamental particle in its
        own right, but rather may represent the collective
        effect of degrees of freedom
        at high energy which have been integrated out of the theory.
        In any such effective field theory
        it is difficult to maintain light scalar
        masses because quantum corrections will
        typically renormalize these to the scale of the cutoff unless
    	they are protected by a symmetry.
        This difficulty afflicts all particulate theories of dark energy
        equally, and we have nothing new to contribute to this debate.
        
        The coupling functions $B$ and $B_H$ are unknown, although they
        will be subject to certain restrictions if we wish the dark energy
        field to exhibit an acceptable chameleon phenomenology.
        We will not impose any such restrictions,
        except to observe that the coupling functions for the $W^\pm$,
        $Z$ and $\gamma$ kinetic terms
        must be the same if Eq.~\eref{eq:action} is to descend from
        an unbroken gauge-invariant theory of $SU(2) \times U(1)$
        at higher energies.
        Moreover, the coupling functions multiplying the mass terms
        must be the same if we suppose that the $W^\pm$ and $Z$ obtain
        their masses via spontaneous symmetry breaking, and that the
        Higgs sector consists of a minimal $SU(2)$ doublet.
        Since we wish to retain both
        these phenomenological successes of the Standard Model, we are left
        with \emph{at most} two free coupling functions.
		In many cases, however, we expect that Eq.~\eref{eq:action} will
		not have a UV completion unless these couplings are the same, because
		the longitudinal polarizations of the $Z$ and $W^\pm$
		are associated with Goldstone modes of the Higgs.
        
        \section{Electroweak precision observables}
        \label{sec:ewpo}
                
        \subsection{Constraints from Z decay}
        \label{sec:Z-decay}

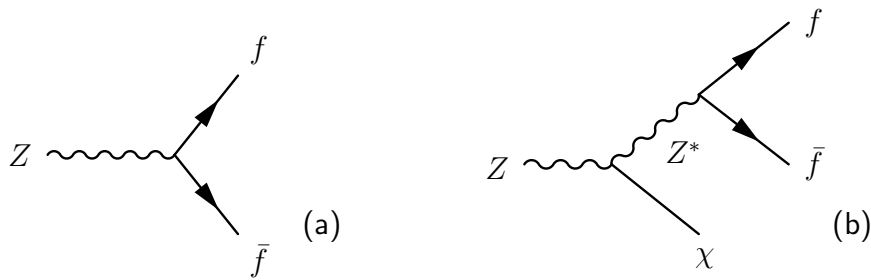
\begin{figure}
                \vspace{3mm}
                \hfill
                \begin{fmfgraph*}(80,60)
                        \fmfleft{l}
                        \fmfright{r1,r2}
                        \fmf{boson}{l,v}
                        \fmf{fermion}{v,r1}
                        \fmf{fermion}{v,r2}
                        \fmfv{label=$Z$}{l}
                        \fmfv{label=$f$}{r2}
                        \fmfv{label=$\bar{f}$}{r1}
                \end{fmfgraph*}
                \hspace{3mm}
                \figlabel{a}
                \hfill
                \begin{fmfgraph*}(100,80)
                        \fmfleft{l}
                        \fmfbottom{b}
                        \fmfright{r1,r2}
                        \fmf{boson}{l,v1}
                        \fmf{boson,label=$Z^\ast$}{v1,v2}
                        \fmf{plain}{v1,b}
                        \fmf{fermion}{v2,r1}
                        \fmf{fermion}{v2,r2}
                        \fmfv{label=$Z$}{l}
                        \fmfv{label=$\chi$}{b}
                        \fmfv{label=$f$,label.angle=0}{r1}
                        \fmfv{label=$\bar{f}$,label.angle=0}{r2}
                        \fmffreeze
                        \fmfforce{(0.33w,0.33h)}{v1}
                        \fmfforce{(0.66w,0.66h)}{v2}
                        \fmfforce{(w,h)}{r1}
                        \fmfforce{(w,0.33h)}{r2}
                        \fmfforce{(0,0.33h)}{l}
                        \fmfforce{(0.66w,0)}{b}
                \end{fmfgraph*}
                \hspace{3mm}
                \figlabel{b}
                \hfill
                \mbox{} \\[3mm]
                \caption{\label{fig:z-decay} Contributions to the decay width of
                the neutral $Z$ boson. In \figlabel{a}, an on-shell $Z$ decays
                to a fermion--antifermion pair $f \bar{f}$.
                In \figlabel{b}, the decay is precipitated by emission of
                a dark energy particle, $\chi$, leaving the original
                $Z$ in an off-shell excited state which subsequently decays
                to $f \bar{f}$. If the final-state $\chi$ is not observed, these
                processes cannot be distinguished and therefore both contribute
                to the decay width into $f \bar{f}$.}
        \end{figure}

        Let us first consider corrections where some dark
        energy quanta are present in the final state.
        These corrections can be considered as a form
        of ``dark energy bremsstrahlung''.
        Since the final-state dark energy particles
		escape the detector and are not observed,
        such reactions look like extra contributions to the cross-section
        for the corresponding bare Standard Model process.
        Among the best-measured
        of these is the width for $Z$ decay,
        depicted
        for decay into a fermion--antifermion pair $f \bar{f}$
        of common mass $M_f$
        with and without dark energy dressing
        in Figs.~\ref{fig:z-decay}\figlabel{a} and \figlabel{b}
        respectively.
        In the dressed process \figlabel{b} the
        on-shell 4-momentum of one outgoing fermion
        (which we label `2' by convention)
        is fixed by conservation of 3-momentum.
        The energy of the other fermion is
        determined by energy conservation in terms of a quadratic
        equation to be given below.
        We show in \ref{sec:z-calculation} that the differential
        contribution to the Z decay width
        from emission of a single dark energy particle
        of energy $E_\chi$ into a solid angle $\d\Omega_\chi$
        satisfies
        \begin{equation}
                \fl
                \frac{\d \Gamma(Z \rightarrow \chi f \bar{f})}
                     {\Gamma(Z \rightarrow f \bar{f})}
                =
                \frac{\bar{B}^{\prime 2}}{(2\pi)^3}
                \frac{M_Z^2}{M^2}
                \d \hat{E}_\chi \, \d \Omega_\chi \;
                \frac{\sqrt{\hat{E}_\chi^2 - y^2}\sqrt{\hat{E}_1^2 - x^2}}
                        {J (1 - \hat{E}_\chi - \hat{E}_1)(1 + \hat{r}^2)^2}
                \frac{\mathcal{M}_{\chi f \bar{f}}}{\mathcal{M}_{f \bar{f}}} ,
                \label{eq:rate-ratio}
        \end{equation}
        where $M = \beta^{-1}$ is the dark energy coupling scale, and
        $x$ and $y$ are defined by
        \begin{eqnarray}
                x & \equiv \frac{M_f^2}{M_Z^2} \label{eq:x-def} \\
                y & \equiv \frac{M_\chi^2}{M_Z^2} \label{eq:y-def} .
        \end{eqnarray}
        The outgoing dark energy scalar is taken to have 3-momentum
        $\vect{q}$.
        We introduce dimensionless ``hatted'' energies and momentum
        according to the rules
        \begin{eqnarray}
                \hat{E}_i = \frac{E_i}{M_Z} \label{eq:ehat-def} \\
                \hat{\vect{q}} = \frac{\vect{q}}{M_Z} \label{eq:qhat-def} .
        \end{eqnarray}
        where $i \in \{ \chi, 1, 2 \}$.
        The quantity $\hat{r}^2$ measures the degree to which the intermediate
        $Z^\ast$ is off-shell, and satisfies
        \begin{equation}
                \hat{r}^2 \equiv -1 + 2 \hat{E}_\chi - y^2 .
        \end{equation}
        It is equal to $-1$ for an intermediate $Z$ which is precisely
		on-shell, although in this limit the finite width of the
		$Z$ cannot be ignored.
        The energy $\hat{E}_1$ must be a solution of the quadratic
        equation
        \begin{eqnarray}
                \fl\nonumber
                \hat{E}_1^2 \left\{
                        \cos^2 \theta ( \hat{E}_\chi^2 - y^2 ) - (1 - \hat{E}_\chi^2 )
                \right\}
                + \hat{E}_1 ( 1 - \hat{E}_\chi )( 1 + y^2 - 2 \hat{E}_\chi) \\
                = \frac{1}{4}(1 + y^2 - 2 \hat{E}_\chi)^2 +
                x^2(\cos^2 \theta)(\hat{E}_\chi^2 - y^2) ,
                \label{eq:E-equation}
        \end{eqnarray}
        where $\theta$ is the angle between $\vect{q}$ and the 3-momentum
        of fermion 1. Although two solutions for $\hat{E}_1$ exist,
        one is always spurious. The solutions change roles at
        $\theta = \pi/2$.
        Moreover, $J$ is a Jacobian arising from fixing $\hat{E}_1$ to be
        a solution of Eq.~\eref{eq:E-equation}.
        It is defined by
        \begin{equation}
                J = \left| 1 + \hat{E}_1
                        \frac{1 + (\hat{E}_\chi^2 - y^2)^{1/2}(\hat{E}_1^2 - x^2)^{-1/2}
                                \cos \theta}
                                {1 - \hat{E}_\chi - \hat{E}_1}
                        \right| .
                \label{eq:jacobian-j}
        \end{equation}
        The matrix element $\mathcal{M}_{f \bar{f}}$
        satisfies
        \begin{equation}
                \mathcal{M}_{f \bar{f}}
                =
                \sqrt{1 - 4 x^2}
                \left\{ 6 g_L g_R x^2 + (g_L^2 + g_R^2)(1-x^2) \right\} ,
                \label{eq:direct-decay-matrix}
        \end{equation}
        where $g_L$ and $g_R$ are the left- and right-handed couplings
        of the fermion species to the $Z$;
        and
        $\mathcal{M}_{\chi f \bar{f}}$ is a complicated function whose
        form is determined in \ref{sec:z-calculation}
		and which can be read off from
		Eqs.~\eref{eq:accompanied-rate}--\eref{eq:accompanied-decay}
		or Eqs.~\eref{eq:A-def}--\eref{eq:B-def}.
        After integrating over $\hat{E}_\chi$ and
        the solid angle $\Omega_\chi$,
        Eq.~\eref{eq:rate-ratio}
        determines the cross-section for any dressed process in terms of the
        bare standard model cross-section.
        In a generic model without fine-tuning, for which
        $\bar{B}' \sim 1$, this rate takes the
        form
        \begin{equation}
                \frac{\Gamma(Z \rightarrow \chi f \bar{f})}
                     {\Gamma(Z \rightarrow f \bar{f})}
                =
                \frac{1}{16 \pi^3}
                \frac{M_Z^2}{M^2} I_{\chi f \bar{f}} ,
                \label{eq:total-ratio}
        \end{equation}
        and $I_{\chi f \bar{f}}$ is found to be numerically
        of order $I_{\chi f \bar{f}} \approx 0.2$ for a wide
        range of fermion masses and couplings.
        The width of the $Z$ into visible particles is predicted to be
        $\Gamma_Z = 2.4952$~GeV within the Standard Model, with a small
        theoretical error.
        Its measured value is $\Gamma_Z = (2.4952 \pm 0.0023)$~GeV
        \cite{Amsler:2008zz},
        implying that any enhancement due to dark energy will be compatible
        with observation only if $M \gtrsim 0.66 M_Z \sim 60$~GeV.
		Moreover, our neglect of the $Z$ width means that this
		is a conservative over-estimate.
        Thus, under the very mild constraint
        $M \gtrsim M_Z$ it seems clear that there will be no disagreement
        with the data. Processes similar to
		Fig.~\ref{fig:z-decay}\figlabel{b}, but with emission of more than
		one dark energy particle into the final state, are suppressed
		by extra powers of $(M_Z/M)^2(2\pi)^{-3}$.
        
        Dark energy bremsstrahlung could have consequences beyond
        enhancements to decay widths and cross-sections of the
        sort calculated above.
        Soft bremsstrahlung effects could be significant
        in QCD if they initiated
        jet formation by destabilizing
        quarks or gluons, or if their aggregate effect could be resolved
        by partons participating in a sufficiently hard collision.
        However, such effects are likely to be important only if the
        dark energy couples at a very low scale.
        We can estimate that the S-matrix element for
        any bremsstrahlung event should controlled by
        the square of the single-chameleon coupling constant,
        of order $M_f/M$ for a fermion of mass $M_f$,
        and a phase space factor of order $\ln s/M_\chi^2$, where $s \approx
        \Mew^2$ is the usual Mandelstam variable
        and $M_\chi \lesssim 10^{-8} \; \mbox{eV}$ is the
        dark energy mass in the beam pipe
        \cite{Donnachie:2002en,Dissertori:2003pj}.
        The logarithm is roughly of order $10^2$.
        A significant effect can occur if the product
        $(M_f/M)^2 \ln \Mew^2/M_\chi^2 \sim 1$,
        but unless the dark energy scalar is very light
        this combination is generally negligible whenever the coupling
        scale $M$ is modestly larger than the mass
        of the fermion species in question,
        of order $M \gtrsim 10^2 M_f$.

        \subsection{Oblique corrections}
        \label{sec:oblique}
                
\begin{figure}
                \vspace{3mm}
                \hfill
                \begin{fmfgraph*}(100,60)
                        \fmfleft{l1,l2}
                        \fmfright{r1,r2}
                        \fmf{fermion}{l1,v1}
                        \fmf{fermion}{l2,v1}
                        \fmf{boson}{v1,v2}
                        \fmf{fermion}{v2,r1}
                        \fmf{fermion}{v2,r2}
                \end{fmfgraph*}
                \figlabel{a}
                \hfill
                \mbox{}
                \\[8mm]
                \mbox{}
                \hfill
                \begin{fmfgraph*}(100,60)
                        \fmfleft{l1,l2}
                        \fmfright{r1,r2}
                        \fmf{fermion}{l1,v1}
                        \fmf{fermion}{l2,v1}
                        \fmf{boson}{v1,v2}
                        \fmf{fermion}{v2,r1}
                        \fmf{fermion}{v2,r2}
                        \fmffreeze
                        \fmf{plain,left=30,tension=1.4}{v1,v1}
                        \fmf{plain,left=150,tension=1.4}{v1,v1}
                        \fmf{plain,left=-30,tension=1.4}{v1,v1}
                        \fmf{plain,left=150,tension=1.4}{v2,v2}
                        \fmf{plain,left=-150,tension=1.4}{v2,v2}
                \end{fmfgraph*}
                \figlabel{b}
                \hspace{3mm}
                \hfill
                \hspace{3mm}
                \begin{fmfgraph*}(100,60)
                        \fmfleft{l1,l2}
                        \fmfright{r1,r2}
                        \fmf{fermion}{l1,v1}
                        \fmf{fermion}{l2,v1}
                        \fmf{boson}{v1,v2}
                        \fmf{fermion}{v2,r1}
                        \fmf{fermion}{v2,r2}
                        \fmffreeze
                        \fmf{plain,left=0.5}{v1,v2}
                        \fmf{plain,left=1.0}{v1,v2}
                        \fmf{plain,left=1.5}{v1,v2}
                \end{fmfgraph*}
                \figlabel{c}
                \hspace{3mm}
                \hfill
                \mbox{}
                \\[8mm]
                \mbox{}
                \hfill
                \begin{fmfgraph*}(100,60)
                        \fmfleft{l1,l2}
                        \fmfright{r1,r2}
                        \fmf{fermion}{l1,v1}
                        \fmf{fermion}{l2,v1}
                        \fmf{boson,tension=1.5}{v1,v2}
                        \fmf{boson,right,tension=0.5}{v2,v3}
                        \fmf{plain,left,tension=0.5}{v2,v3}
                        \fmf{boson,tension=1.5}{v3,v4}
                        \fmf{fermion}{v4,r1}
                        \fmf{fermion}{v4,r2}
                \end{fmfgraph*}
                \figlabel{d}
                \hspace{3mm}
                \hfill
                \mbox{}
                \caption{\label{fig:dark-energy-processes}Classes of dark
                energy diagrams associated with Standard Model reactions,
                exemplified in the case of $2 \rightarrow 2'$ fermion
				scattering.
                Solid lines with arrows represent fermions; wavy lines 
				represent
                the gauge bosons of the electroweak force; and plain lines
                represent dark energy particles.
                The bare Standard Model process is given in \figlabel{a}.
                In \figlabel{b}, the vertices of the reaction are
                dressed by \emph{daisies} which begin and end at the same
                vertex.
                In \figlabel{c}, dark energy quanta \emph{bridge}
                between two different vertices.
                Corrections such as \figlabel{b}--\figlabel{c} which depend
                on the process under study (in this case, depending
                on the initial and final fermion species, and the
                identity of the exchanged boson) are called \emph{straight}.
                On the other hand,
                corrections such as \figlabel{d} which are universal
                for all processes involving the exchange of a given species
                of vector boson are called \emph{oblique}.
                (In principle there are also oblique corrections to the
                fermion species, but typically these do not contribute
                significantly to observable quantities.)
                In general, the dark energy correction to
                \figlabel{a} consists of summing over all possible combinations
                of processes similar to \figlabel{b}--\figlabel{d}.}
        \end{figure}
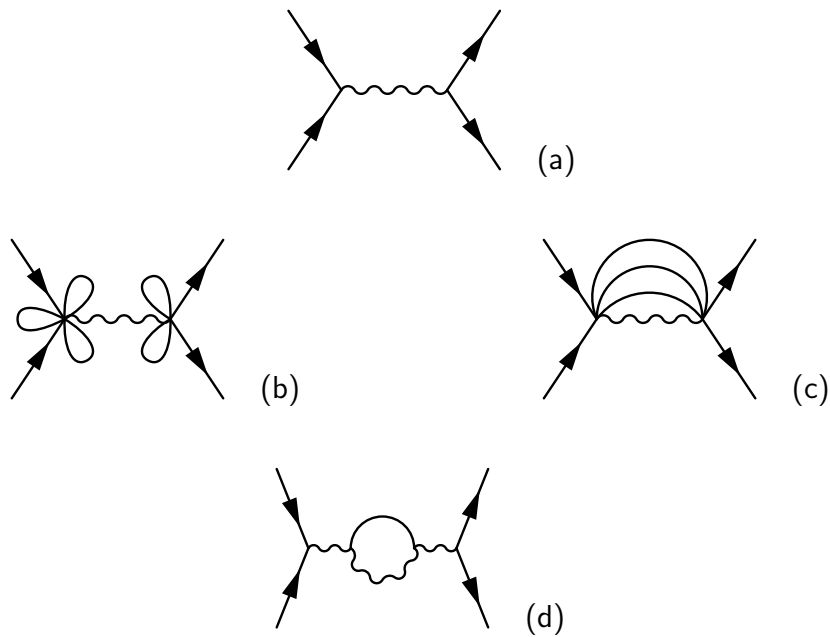

		In addition to bremsstrahlung processes,
        the perturbation theory constructed from
        Eq.~\eref{eq:action} describes processes by which
        Standard Model particles may radiate into an intermediate state
        containing an arbitrary number of dark energy quanta.
        If we exclude reactions in which dark energy particles are present
        in the initial or final state
        then all such processes
        are built out of interactions which are already
        present in the bare Standard Model.
        To study them
        we should begin with a given Standard Model reaction,
        exemplified
        for the case of $2 \rightarrow 2'$ scattering of light fermions
        in Fig.~\ref{fig:dark-energy-processes}\figlabel{a},
        and account for the effect of dark energy activity. This activity
        can naturally be divided into three categories, corresponding to
        Figs.~\ref{fig:dark-energy-processes}\figlabel{b}--\figlabel{d}.

        In Fig.~\ref{fig:dark-energy-processes}\figlabel{b},
        dark energy loops dress each vertex in the
        bare reaction with so-called
        \emph{daisies}, whereas in
        Fig.~\ref{fig:dark-energy-processes}\figlabel{c}
        dark energy quanta \emph{bridge} between two different vertices.
        More complicated bridges, including internal vertices which
        may themselves be dressed by daisies, can also be constructed.%
                \footnote{Note, however, that
                we do not include loops in which the chameleon interacts
                with itself:
                as has been said,
                these are assumed to have been absorbed in the
                parameters of the dark energy model.}
        Together with processes where one or more dark energy quantum
        appears in the final state,
        these are examples of so-called \emph{straight} corrections
        which depend on the process which under consideration
        \cite{Peskin:1990zt}.

        In contradistinction,
        Fig.~\ref{fig:dark-energy-processes}\figlabel{d}
        represents an example of an \emph{oblique} correction,
        which involves intermediate dark
        energy states only in the interior of a gauge boson
        propagator.
        Once an oblique correction has been calculated for a given
        species of gauge field, it is universal for all processes involving
        exchange of that boson.
        In principle, these corrections are all equally important
        and for a general momentum transfer $q$
        it is a complicated process to compute them.
        However, we argue in \ref{sec:bridges} that the daisies and bridges
        which constitute the straight corrections
        are momentum-independent up to terms of order $q^2 / M^2$, where
        $M > \Mew$ is a dark energy coupling scale
        characteristic of the fermion species which participate.
        Provided they are the same for all species,
        such momentum-independent terms can be absorbed in a renormalization
        of the Fermi constant, $G_F$, and are therefore unobservable.
        We have seen in \S\ref{sec:Z-decay} that
		in any phenomenologically acceptable scenario we expect
        $M \gg \Mew$, implying that the
        remaining contributions can be neglected
        in comparison with that of the oblique correction,
        Fig.~\ref{fig:dark-energy-processes}\figlabel{d},
        which is present at order $q^2 / \Mew^2$.
 		Oblique corrections
        will therefore
		give the most stringent constraints if they turn out
        to require $M \gtrsim 100$~GeV.

        The effect of physics beyond the Standard Model has been studied by
        many authors,
        and is frequently dominated by oblique corrections.
        Peskin \& Takeuchi \cite{Peskin:1990zt,Peskin:1991sw}
        introduced a simple parametrization of them
        in terms of three quantities
        $S$, $T$ and $U$ which quantify the magnitude of
        corrections near zero momentum transfer,%
                \footnote{An alternative parametrization was proposed
				simultaneously by Altarelli \& Barbieri
                \cite{Altarelli:1990zd,Altarelli:1991fk}.}
        but assumed that whatever new physics was responsible
        for modifying the properties of the
        gauge bosons was heavy.
        This assumption was later removed by
        Maksymyk, London \& Burgess
        \cite{Maksymyk:1993zm,Burgess:1993mg},
        who introduced
        new parameters $V$, $W$ and $X$ to quantify the significance
        of radiative corrections around the $Z$ resonance.%
                \footnote{See also Refs.~\cite{Kundu:1996ah,Rosner:1993rj}.}
        In the remainder of
        this section, we briefly review the parametrization of oblique
        corrections in terms of $S$, $T$, $U$, $V$, $W$ and $X$.
        
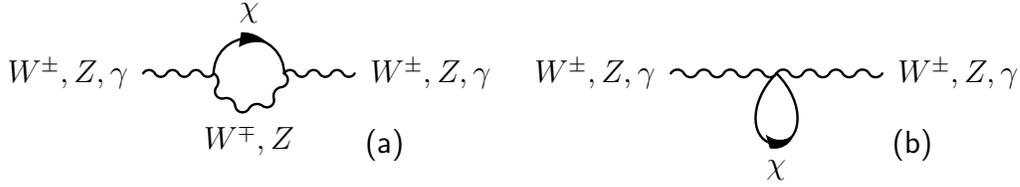
\begin{figure}
                \begin{center}
                        \hfill
                        \begin{fmfgraph*}(80,60)
                                \fmfleft{l}
                                \fmfright{r}
                                \fmfpen{thin}
                                \fmf{boson,tension=2}{l,v1}
                                \fmf{fermion,label=$\chi$,left}{v1,v2}
                                \fmf{boson,label=$W^\mp,, Z$,left}{v2,v1}
                                \fmf{boson,tension=2}{v2,r}
                                \fmfv{label=$W^\pm,, Z,, \gamma$}{l}
                                \fmfv{label=$W^\pm,, Z,, \gamma$}{r}
                        \end{fmfgraph*}
                        \figlabel{a}
                        \hspace{3mm}
                        \hfill
                        \hspace{3mm}
                        \begin{fmfgraph*}(80,60)
                                \fmfleft{l}
                                \fmfright{r}
                                \fmfpen{thin}
                                \fmf{boson,tension=2}{l,v}
                                \fmf{fermion,label=$\chi$,left}{v,v}
                                \fmf{boson,tension=2}{v,r}
                                \fmfv{label=$W^\pm,, Z,, \gamma$}{l}
                                \fmfv{label=$W^\pm,, Z,, \gamma$}{r}
                        \end{fmfgraph*}
                        \figlabel{b}
                        \hfill
                        \mbox{}
                \end{center}
                \caption{\label{fig:polarizations}Processes contributing to the
                self-energy of the intermediate vector bosons
                $\gamma$, $W^\pm$ and $Z$.
                An initial vector boson state, represented by a wavy line,
                radiates into scalar quanta $\chi$ (represented by a solid line)
                which are eventually re-absorbed
                to yield a final state characterized by the same quantum numbers
                and momentum as the initial state.}
        \end{figure}

        The one-loop obliquely-corrected
        vector boson propagators are obtained by summing over an
        arbitrary number of insertions of the one-loop diagrams of
        Fig.~\ref{fig:polarizations} in the tree-level propagator.
        In unitarity gauge, where the three
        would-be Goldstone modes supplied by the Higgs doublet have been
        absorbed as longitudinal polarizations of the $W^\pm$ and $Z$,
        the tree-level propagator for each massive vector
        boson can be written
        \begin{equation}
                \langle X^a_A(k_1) X^{\dag b}_B(k_2) \rangle
                = -\im (2\pi)^3 \delta(k_1 + k_2) \delta_{AB}
                  \left( \eta^{ab} + \frac{k^a k^b}{m^2_A} \right)
                  \Delta(k^2) ,
                \label{eq:propagator-matrix}
        \end{equation}
        where we have defined $k \equiv k_1 = - k_2$
        and
        the quantum field $X_A$ is built out of the creation and
        annihilation operators corresponding to a vector boson of species $A$
        and mass $m_A$.
        The tree-level propagator function
        satisfies $\Delta^{-1}(k^2) = k^2 + m_A^2$.
        The photon propagator can be written in an analogous form,
        with $m_A \mapsto 0$ in the function $\Delta(k^2)$ and
        $m_A^2 \mapsto -k^2$ in the tensor prefactor.
        
        We define the sum of the one-particle-irreducible diagrams which
        connect an initial-state vector boson of species $A$ with a
        final-state vector boson of species $B$ and carrying momentum
        $k$ to be $\im \Pi_{AB}^{ab}(k^2)/(2\pi)^4$. Since the $Z$ and $\gamma$
        are electrically neutral they are permitted to mix beyond tree-level,
        which would
        correspond to a non-zero vacuum polarization $\Pi_{Z\gamma}^{ab}$.
        However, inspection of the interactions in Eq.~\eref{eq:action}
        shows that
        Eq.~\eref{eq:action} does not induce extra mixing
        and we can set $\Pi_{Z\gamma}^{ab} = 0$.

        With this simplification, the full propagator can be resummed
        using the Schwinger--Dyson equations. The result is that
        the propagator function $\Delta$ in Eq.~\eref{eq:propagator-matrix}
        should be replaced by a resummed function $\Delta'$, which
        for each species $A$ satisfies
        \begin{equation}
                \Delta'(k^2) = \frac{1}{k^2 + m_A^2 - \Pi^{(0)}_{AA}(k^2)} ,
                \label{eq:resummed-propagator}
        \end{equation}
        where we have written
        \begin{equation}
                \Pi^{ab}_{AB}(k^2) =
                  \eta^{ab} \Pi^{(0)}_{AB}(k^2) + k^a k^b \Pi^{(2)}_{AB}(k^2) ,
        \end{equation}
        and, for external states which consist only of light fermions
        of invariant mass-squared $M_f^2$,
        Eq.~\eref{eq:resummed-propagator} is valid up to corrections
        of order $M_f^2/M_W^2$ which we neglect.
        Therefore,
        the quadratic term $\Pi^{(2)}_{AB}$ will not appear in the
        remainder of this paper, and to simplify notation we write all
        subsequent formulae in terms of the abbreviation
        $\Pi_{AB} \equiv \Pi^{(0)}_{AB}$.
        
        \subsection{The $S$, $T$, $U$, $V$ and $W$ parameters}
        
        In the absence of radiative corrections, the Standard Model
        entails the existence of simple relationships among the observables
        of the theory. Since there are three
        free quantities which parametrize the broken phase%
        ---the two gauge couplings $g_1$ and $g_2$, together with the
        Higgs vacuum expectation value---%
        it is necessary to take three masses or couplings from experiment.
        Once this so-called `input parameter set' has been selected, all
        other observables can be expressed in terms of the chosen three.
        In the electroweak sector it is conventional to choose the
        input parameter set to comprise the
        fine structure constant $\alpha$, the Fermi coupling $G_F$,
        and the $Z$ mass, $M_Z$, which are presently the best measured
        electroweak quantities.
        
        With the inclusion of radiative corrections, the
        original simple relationships
        among observables are modified. Indeed,
        in order to match the precision with
        which accelerator experiments can measure electroweak parameters,
        it is usually necessary to include several orders of radiative
        corrections which arise purely within the Standard Model.
        It may happen that these corrections are insufficient to account for
        the deviation of all observables from their tree-level values.
		The remainder must be ascribed to new physics: it is
        only
        this contribution from new physics which we wish to attribute to
        the effect of a dark energy scalar species.
        The relevant
        observables other than $\{ \alpha, G_F,
        M_Z \}$ are the mixing angle, $\thetaw$, and the $W^\pm$ mass, $M_W$,
        together with any cross-sections or decay rates
        which can be written in terms of all these quantities.
        At tree level, $\thetaw$ and $M_W$
        are related to the input parameter set via the rules
        \begin{eqnarray}
                \sin^2 \thetaw (1 - \sin^2 \thetaw) = 
                \frac{\alpha}{16 \sqrt{2} \pi M_Z^2 G_F} ,
                \label{eq:weak-angle} \\
                M_W^2 = M_Z^2 \cos^2 \thetaw ,
                \label{eq:w-mass}
        \end{eqnarray}
        where $\cos \thetaw$ in Eq.~\eref{eq:w-mass} is to be computed from the
        solution to Eq.~\eref{eq:weak-angle}.
        
        The physical mass of a single-particle state corresponding to a
        vector boson is given by the pole of Eq.~\eref{eq:resummed-propagator},
        which renormalizes the Lagrangian parameter $m_A$. Therefore,
        the physical mass $M_A$ satisfies
        \begin{equation}
                M_A^2 = \tilde{M}_A^2
                        \left( 1 - \frac{\Pi_{AA}(-M_A^2)}{M_A^2} \right) ,
                \label{eq:mass-shift}
        \end{equation}
        where we have introduced a useful notation in which
        a tilde, as in $\tilde{M}_A$, denotes the value taken by a quantity
        in the Standard Model without oblique corrections.
        At tree-level, $\tilde{M}_A^2$ is simply equal to $m_A^2$,
        but Eq.~\eref{eq:mass-shift} continues to apply
        \emph{to leading order in radiative corrections}
        even if we allow the vector boson masses
        to receive renormalizations from loops purely within the
        Standard Model.
        On the other hand, the Fermi constant $G_F$ is defined as the
        coupling of the charged-current interaction at zero momentum transfer
        and receives an oblique correction \cite{Peskin:1990zt,Peskin:1991sw}
        \begin{equation}
                G_F = \tilde{G}_F
                        \left( 1 + \frac{\Pi_{WW}(0)}{M_W^2} \right) .
                \label{eq:fermi-shift}
        \end{equation}
        Likewise, the fine structure constant measures the
        electromagnetic interaction at zero momentum transfer
        and receives an oblique correction
        from the photon self-energy,
        \begin{equation}
                \alpha = \tilde{\alpha}
                        \left( 1 + \hat{\Pi}_{\gamma\gamma}(0) \right) ,
                \label{eq:charge-shift}
        \end{equation}
        where $\hat{\Pi}_{\gamma\gamma}(k^2) \equiv
        \Pi_{\gamma\gamma}(k^2)/k^2$.
        Eqs.~\eref{eq:fermi-shift} and~\eref{eq:charge-shift} apply
        even if we allow $\tilde{G}_F$ and $\tilde{\alpha}$ to receive 
		corrections
        from pure Standard Model loops.
        It follows that we can write
        \begin{equation}
                \frac{\sw^2}{\tildesw^2} =
                1 + \frac{\alpha}{4 \sw^2 (\cw^2 - \sw^2)} S
                  - \frac{\alpha \cw^2}{\cw^2 - \sw^2} T
        \end{equation}
        and
        \begin{equation}
                \frac{M_W^2}{\tilde{M}_W^2} =
                1 - \frac{\alpha}{2 (\cw^2 - \sw^2)} S
                  + \frac{\alpha \cw^2}{\cw^2 - \sw^2} T
                  + \frac{\alpha}{4 \sw^2} U ,
        \end{equation}
        where we have introduced the useful abbreviations
        $\sw \equiv \sin \thetaw$ and $\cw \equiv \cos \thetaw$,
        and the parameters $S$, $T$ and $U$ are defined by
        \cite{Peskin:1990zt,Peskin:1991sw,Maksymyk:1993zm}%
                \footnote{Certain terms in these expressions change sign
                depending on the choice of signature for the metric.
                Under reverse of sign convention
                (which gives the timelike convention widely used in
                particle physics, in comparison with the spacelike
                convention adopted in this paper), the formulae
                for $S$, $T$ and $U$ should be modified by reversing the sign 
				of
                each mass-square $M_A^2$, together with extra
                signs for each explicit factor of $k^2$ or $\d k^2$.
                This explains the difference in signs between
                Eqs.~\eref{eq:S}--\eref{eq:U} and the original references,
                which used the signature $(+,-,-,-)$.
                Note also that
                in theories where the $Z$ and $\gamma$ mix beyond
                tree-level,
                $S$ and $U$ receive extra contributions. For details,
                see Refs.~\cite{Peskin:1990zt,Peskin:1991sw,Maksymyk:1993zm}.}
        \begin{eqnarray}
                \frac{\alpha}{4 \sw^2 \cw^2} S \equiv
                        \frac{\Pi_{ZZ}(0) - \Pi_{ZZ}(-M_Z^2)}{M_Z^2}
                        - \hat{\Pi}_{\gamma\gamma}(0)
                        \label{eq:S} , \\
                \alpha T \equiv
                        \frac{\Pi_{ZZ}(0)}{M_Z^2} - \frac{\Pi_{WW}(0)}{M_W^2}
                        \label{eq:T} , \\
                \frac{\alpha}{4 \sw^2} ( U + S ) \equiv
                        \frac{\Pi_{WW}(0) - \Pi_{WW}(-M_W^2)}{M_W^2}
                        - \hat{\Pi}_{\gamma\gamma}(0)
                        \label{eq:U} .
        \end{eqnarray}
        Experimentally observable
        quantities such as the Veltman $\rho$-parameter,
        Eq.~\eref{eq:veltman-rho}, can be written in terms of $S$, $T$
        and $U$.
        
        Electroweak data is not limited to measurements of the $W^{\pm}$ and
        $Z$ masses and the mixing angle, but includes cross-sections and
        decay rates. The standard LSZ formula \cite{Lehmann:1954rq} implies
		that
        the first-order shifts from oblique corrections in these
        quantities can be obtained from their tree-level values
        together with appropriate multiplication by
        wave function renormalization factors $\wavefn_A$, defined for each
        species of massive boson $A$ by the rule
        \begin{equation}
                \wavefn_A \equiv 1 + \left. \frac{\d}{\d k^2} \Pi_{AA}(k^2)
                        \right|_{k^2 = - M_A^2} .
        \end{equation}
        To take account of these factors, it is necessary to introduce two
        further parameters $V$ and $W$
        \cite{Maksymyk:1993zm,Kundu:1996ah,Rosner:1993rj}%
                \footnote{In theories where the $Z$ and $\gamma$ can mix
                beyond tree-level, it is necessary to introduce a third
                new parameter, $X$. See Ref.~\cite{Maksymyk:1993zm}.}
        \begin{eqnarray}
                \alpha V & \equiv
                        \left. \frac{\d}{\d k^2} \Pi_{ZZ}(k^2) \right|_{k^2 = -M_Z^2}
                        - \frac{\Pi_{ZZ}(0) - \Pi_{ZZ}(-M_Z^2)}{M_Z^2} , \\
                \alpha W & \equiv
                        \left. \frac{\d}{\d k^2} \Pi_{WW}(k^2) \right|_{k^2 = -M_W^2}
                        - \frac{\Pi_{WW}(0) - \Pi_{WW}(-M_W^2)}{M_W^2} .
        \end{eqnarray}
        Oblique dark energy corrections to all
        purely electroweak observables can be written in terms of
        $S$, $T$, $U$, $V$ and $W$.
        
        These parameters have simple physical interpretations.
        $S$ is a measure of the difference between the
        wavefunction renormalization of the $Z$ boson and the photon, $\gamma$.
        In an interacting theory, a state
        prepared with definite particle content and momentum
        at some early time may not manifest the same content when probed at
        a later time because the particles may radiate into any other states
        to which they couple. The probability for this to occur is
        quantified by the wavefunction renormalization.
        
        $T$ is a measure of the extra isospin breaking at zero momentum
        which is contributed by new physics. This difference manifests
        itself in the
        relative strength of the charged- and neutral-current interactions.
        The precise
        balance between these interactions may be upset by coupling
        to the dark energy scalar, but in the Standard Model with
        a minimal Higgs sector $T$ is unlikely
        to receive large corrections
        unless isospin symmetry is broken explicitly at
        tree level.
        Similarly, $U$ is a measure
        of the difference between the $W^{\pm}$ and $Z$ wavefunction
        renormalizations.
        Finally,
        $V$ and $W$ quantify the difference between the wavefunction
        renormalizations of the $Z$ and $W^{\pm}$ bosons, respectively,
        on the mass-shell, compared with zero momentum.
        In what follows, we will see this structure emerge explicitly from
        our analysis.
        
        \section{Vector boson vacuum polarizations}
        \label{sec:polarization}
        
        To evaluate the $S$, $T$, $U$, $V$ and $W$ parameters, one requires
        an explicit expression for the vector boson vacuum polarizations.
        In this section, we obtain the necessary
        self-energies by calculating the two one-loop diagrams in
        Fig.~\ref{fig:polarizations}.
        
        \subsection{Feynman rules}
        
        Our detailed information concerning the
        properties of the $W^{\pm}$ and $Z$ bosons comes mostly from the
        LEPII experiment, which created these particles abundantly
        in head-on
        $e^+e^-$ collisions. The $W^{\pm}$s and $Z$s synthesized in
        this way were produced at
        rest in the beam-pipe and spent their entire lifetime within its 
		vacuum,
        before decaying into other particles which could subsequently be
        detected.
        In the environment of the beam-pipe, we can assume
        that the scalar field has a constant vacuum expectation value
        $\bar{\chi}$ together with small excitations $\delta\chi$.
        To obtain the one-loop vacuum polarization, it is necessary
        to describe the interactions of the $W^{\pm}$ and $Z$ to order
        $\delta\chi^2$. For interactions involving
        a $W^+$ and $W^-$ the relevant vertices are:
        \vspace{7mm}
        \begin{equation}
                \parbox{30mm}{
                        \begin{fmfgraph*}(60,40)
                                \fmfleft{l1,l2}
                                \fmfright{r}
                                \fmf{photon,label=$W^+_a$,label.side=right}{l1,v}
                                \fmf{photon,label=$W^-_b$,label.side=left}{l2,v}
                                \fmf{plain}{v,r}
                                \fmfv{label=$k_2$}{l1}
                                \fmfv{label=$k_3$}{l2}
                                \fmfv{label=$k_1$}{r}
                        \end{fmfgraph*}
                        }
                        \mapsto \bar{B}' \beta
                                \left[ \eta^{ab} (k_2 \cdot k_3 - \gamma m_W^2 )
                                                - k_2^b k_3^a \right] ,
                \label{feynrule:w-cubic}
        \end{equation}
        \vspace{10mm}
        \begin{equation}
                \parbox{30mm}{
                        \begin{fmfgraph*}(60,40)
                                \fmfleft{l1,l2}
                                \fmfright{r1,r2}
                                \fmf{photon,label=$W^+_a$,label.side=right}{l1,v}
                                \fmf{photon,label=$W^-_b$,label.side=left}{l2,v}
                                \fmf{plain}{r1,v}
                                \fmf{plain}{r2,v}
                                \fmfv{label=$k_2$}{l1}
                                \fmfv{label=$k_3$}{l2}
                                \fmfv{label=$k_1$}{r1}
                                \fmfv{label=$k_4$}{r2}
                        \end{fmfgraph*}
                        }
                        \mapsto \frac{\bar{B}'' \beta^2}{2}
                                \left[ \eta^{ab} (k_2 \cdot k_3 - \epsilon m_W^2 )
                                                - k_2^b k_3^a \right] ,
                \label{feynrule:w-quartic}
        \end{equation}
        \vspace{7mm}
        where $\bar{B}' \equiv B'(\beta \bar{\chi})$,
        $\bar{B}'' \equiv B''(\beta \bar{\chi})$ together with
        equivalent definitions for $B_H$;
        the spacetime inner product is denoted $p \cdot q \equiv p^a q_a$
        for any two four-vectors $p^a$ and $q^a$; and
        we have defined quantities $\gamma$ and $\epsilon$
        according to the rules
        \begin{eqnarray}
                \gamma \equiv \frac{\bar{B}'_H}{\bar{B}'} \frac{\beta_H}{\beta} \\
                \epsilon \equiv \frac{\bar{B}''_H}{\bar{B}''}
                        \frac{\beta_H^2}{\beta^2} .
        \end{eqnarray}
        
        With this choice of Feynman rules, the diagram of
        Fig.~\ref{fig:polarizations}\figlabel{a}
        corresponds to a vacuum polarization of the form
        \begin{equation}
                \fl
                \Pi_{WW}(k^2) = \frac{\beta^2}{8 \pi^2}
                        \frac{\bar{B}^{\prime 2}}{\bar{B}}
                        \int_0^1 \d x \int_0^\Lambda
                        \frac{\kappa^3 \; \d \kappa}{(\kappa^2 + \Sigma^2)^2}
                        \left[
                                \frac{\kappa^2}{4} ( 2k^2 + \gamma^2 M_W^2 )
                                + ( x k^2 + \gamma M_W^2 )^2
                        \right] ,
                \label{eq:diagram-a}
        \end{equation}
        where $x$ is a Feynman parameter,
        and we have Wick rotated to Euclidean signature
        before replacing the Euclidean volume element
        by $2 \pi^2 \kappa^3 \, \d \kappa$.
        The momentum scale
        $\Lambda$ is a sharp cutoff which regulates the maximum
        Euclidean momentum permitted to circulate in the loop, and
        therefore determines the size of the phase space of scalar states
        to which each $W^{\pm}$ couples.%
			\footnote{Power-law divergences in $\Lambda$, if they exist,
			are likely to violate gauge invariant although
			logarithmic divergences should be physically meaningful.
			Also, loop calculations in unitarity gauge are known to
			overestimate power law divergences in certain circumstances.
			These issues were addressed in
			Refs.~\cite{Burgess:1992va,Burgess:1992gx}.
			In the present case it will turn out that we require only
			the logarithmic terms.
			If any power-law divergences were present, however,
			then it would not be possible to interpret the result as
			a quantitative prediction. Instead---provided such powers were
			compatible with na\"{\i}ve dimensional
			analysis (which excludes the possibility of overestimation)
			and the gauge symmetries of the model---the correct
			interpretation would
			be that the calculation under discussion
			was sensitive to the details of UV physics.}
		Finally, $\Sigma^2$
        is an abbreviation for the quantity
        \begin{equation}
                \Sigma^2 \equiv x(1-x) k^2 + (1-x) M_W^2 + x M_\chi^2 .
                \label{eq:sigma}
        \end{equation}
        In writing Eqs.~\eref{eq:diagram-a}--\eref{eq:sigma} we have freely
        replaced $m_W^2$ by $M_W^2$, since the correction this induces is
        formally of higher order in the loop expansion.
        
        The diagram of Fig.~\ref{fig:polarizations}\figlabel{b}
        gives a somewhat simpler contribution,
        \begin{equation}
                \Pi_{WW}(k^2) = - \frac{\beta^2}{8\pi^2}
                        \frac{\bar{B}^{\prime 2}}{\bar{B}} \int_0^1 \d x
                        \int_0^\Lambda \frac{\kappa^3 \; \d \kappa}{\kappa^2 + M_\chi^2}
                        \frac{\Omega}{2} ( k^2 + \epsilon M_W^2 ) ,
                \label{eq:diagram-b}
        \end{equation}
        where $\Omega$ is a dimensionless combination which measures the
        curvature of the coupling function $B$ in the vacuum,
        \begin{equation}
                \Omega \equiv \frac{\bar{B}'' \bar{B}}{\bar{B}^{\prime 2}} .
                \label{eq:omega-def}
        \end{equation}
        
        We also require the vacuum polarization for the $Z$ boson
        and the photon, $\gamma$.
        However, no further calculation is required since the relevant
        Feynman rules can be obtained from~\eref{feynrule:w-cubic}--%
        \eref{feynrule:w-quartic}, and the necessary vacuum polarizations
        can likewise be obtained from
        Eqs.~\eref{eq:diagram-a}--\eref{eq:diagram-b}.
        Since the $\gamma$ and $Z$ are their own antiparticles, each
        vertex in~\eref{feynrule:w-cubic}--\eref{feynrule:w-quartic}
        acquires a symmetry factor of $1/2$. To obtain the correct
        vacuum polarizations, one makes the replacement $M_W \mapsto M_Z$
        in Eqs.~\eref{eq:diagram-a}--\eref{eq:diagram-b} for the $Z$,
        and $M_W \mapsto 0$ for the photon.
        
        Assembling these terms and carrying out the $\kappa$ integrals,
        it follows that the vacuum polarization for each species of boson
        satisfies
        \begin{eqnarray}
                \fl\nonumber
                \Pi_{AA}(k^2) = \frac{\beta^2}{8\pi^2}
                        \frac{\bar{B}^{\prime 2}}{\bar{B}}
                        \int_0^1 \d x \;\Bigg\{ \frac{2 k^2 + \gamma^2 M_A^2}{4}
                                \left[
                                        \Lambda^2 + \frac{\Lambda^2}{2}
                                                \frac{\Lambda^2}{\Lambda^2 + \Sigma^2}
                                        - \Sigma^2 \ln \left(
                                                1 + \frac{\Lambda^2}{\Sigma^2}
                                        \right)
                                \right]
                                \\ \nonumber
                                \mbox{} +
                                (x k^2 + \gamma M_A^2)^2
                                \left[
                                        - \frac{1}{2} \frac{\Lambda^2}{\Lambda^2 + \Sigma^2}
                                        + \frac{1}{2} \ln \left(
                                                1 + \frac{\Lambda^2}{\Sigma^2}
                                        \right)
                                \right]
                                \\
                                \mbox{} -
                                \frac{\Omega}{2} (k^2 + \epsilon M_A^2)
                                \left[
                                        \frac{\Lambda^2}{2} - \frac{M_\chi^2}{2}
                                        \ln \left(1 + \frac{\Lambda^2}{M_\chi^2} \right)
                                \right]
                        \Bigg\}
                        \label{eq:self-energy}
        \end{eqnarray}
        
        \subsection{Effective Lagrangians for the vacuum polarization}
        \label{sec:eft}
        
        Eq.~\eref{eq:self-energy} is a complicated expression from which it
        is difficult to extract the important qualitative features of
        the oblique corrections. To do better, one can analyze
        $\Pi_{AA}(k^2)$ in terms of an effective Lagrangian which would give
        rise to the same vacuum polarization.
        
        \subtitle{A. Low energy, massive vector bosons.}
        Consider first the limit $|q^2| \ll M_W^2$.
        For each species of massive vector boson
        $A$ one can make the expansions
        \begin{equation}
                \frac{\Lambda^2}{\Lambda^2 + \Sigma^2} =
                \frac{1}{\sigma^2} +
                \frac{1}{\sigma^2} \sum_{n = 1}^\infty (-1)^n \left[
                        \frac{x(1-x)}{\sigma^2} \frac{k^2}{\Lambda^2}
                \right]^n
                \label{eq:cutoff-corrections}
        \end{equation}
        and
        \begin{equation}
                \fl
                \ln \left( 1 + \frac{\Lambda^2}{\Sigma^2} \right) =
                \ln \frac{\sigma^2}{\sigma^2 - 1} +
                \sum_{m = 1}^\infty
                \sum_{n = 1}^\infty
                \frac{(-1)^{m(n+1) + 1}}{m \sigma^{2m}}
                \left[
                        \frac{x(1-x) k^2}{(1-x)M_A^2 + x M_\chi^2}
                \right]^{mn}
                \label{eq:electroweak-corrections}
        \end{equation}
        where we have defined $\sigma^2$ by the rule
        \begin{equation}
                \sigma^2 \equiv 1 + (1-x) \frac{M_A^2}{\Lambda^2}
                        + x \frac{M_\chi^2}{\Lambda^2} .
        \end{equation}
        In particular, $\sigma^2 \approx 1$ whenever the scale of the cutoff,
        $\Lambda$, is much larger than the electroweak scale
        $\Mew \sim M_A$. Eq.~\eref{eq:cutoff-corrections} is an expansion
        in powers of $k^2/\Lambda^2$. In an effective field theory, these
        contributions would come from a tower of non-renormalizable operators
        suppressed by the cutoff scale, although one should remember
        that whenever these operators become important the
        bridge corrections discussed in \ref{sec:bridges}
        will also make a significant contribution.
        On the other hand,
        Eq.~\eref{eq:electroweak-corrections} amounts to an
        expansion in powers of $k^2/M_A^2$.%
                \footnote{The series expansion in 
                Eq.~\eref{eq:electroweak-corrections} can be integrated
                term-by-term in $x$, producing an expansion in powers of $k^2/M_A^2$
                with coefficients which involve hypergeometric functions
                of $M_\chi^2/M_A^2$. When expanded in powers of this ratio
                it is possible that logarithms of $M_\chi^2/M_A^2$ are 
				generated,
                although suppressed by a positive power of $M_\chi^2/M_A^2$.
                It follows that the scale at which this tower of
                non-renormalizable operators becomes significant
                genuinely
                \emph{is} around the electroweak scale, $|k^2| \sim M_A^2$.}
        These contributions would come
        from non-renormalizable operators suppressed only by the electroweak
        scale. As we increase the momentum which is transferred through the
        gauge boson propagator from zero, we expect to see corrections
        enter at the scale $|k^2| \sim M_A^2$, followed by another set of
        corrections at the cutoff.
        
        Collecting these expressions
        one finds an expansion for $\Pi_{AA}(k^2)$, which yields
        \begin{equation}
                \fl
                \Pi_{AA} (k^2) = \frac{g^2}{M^2}
                \left[ M_A^2 \alpha_0 + \alpha_2 k^2 + \alpha_4 k^4 +
                        \Or\Big( \frac{k^2}{\Mew^2} \Big)
                \right] ,
                \hfill
                (|k^2| \ll \Mew^2)
                \hspace{1cm}
                \label{eq:pi-expansion-a}
        \end{equation}
        where $g$ is an effective dimensionless coupling constant defined by
        \begin{equation}
                g^2 \equiv \frac{1}{8\pi^2} \frac{\bar{B}^{\prime 2}}{\bar{B}} ,
        \end{equation}
        the mass scale $M$ is $M \equiv \beta^{-1}$, as before,
        and the coefficients $\alpha_i$, for $i \in \{ 0, 2, 4 \}$, satisfy
        \begin{eqnarray}
                \fl
                \alpha_0 & \equiv
                        \frac{\Lambda^2}{4}
                        \left( \frac{\gamma^2}{2} - \Omega \epsilon \right)
                        + \frac{\gamma^2 M_A^2}{16}
                          \left[ 6 \ln \frac{\Lambda^2}{M_A^2} - 1
                        + \Or\Big( \frac{M_\chi^2}{\Mew^2} \Big)
                        \right] ,
                \label{eq:alpha-zero} \\
                \fl
                \alpha_2 & \equiv
                        \frac{\Lambda^2}{4} ( 1 - \Omega ) +
                        \frac{M_A^2}{144} \left[
                                6 [ \gamma(12 - \gamma) - 6 ]
                                \ln \frac{\Lambda^2}{M_A^2}
                                + \gamma(36 - 5 \gamma) - 18 
                                + \Or\Big( \frac{M_\chi^2}{\Mew^2} \Big)
                        \right] ,
                \label{eq:alpha-two} \\
                \fl
                \alpha_4 & \equiv
                        \frac{1}{12} \ln \frac{\Lambda^2}{M_A^2} + \frac{5}{72}
                        + \Or\Big( \frac{M_\chi^2}{\Mew^2} \Big) .
                \label{eq:alpha-four}
        \end{eqnarray}
        We could equally well have obtained this vacuum polarization
        if we had started from an action of the form
        \begin{eqnarray}
                \fl\nonumber
                S = \frac{1}{2} \int \d^4 x \; \Big[
                        \left( 1 - \frac{g^2}{M^2} \alpha_2 \right)
                                \varphi \partial^2 \varphi
                        - M_A^2 \left( 1 - \frac{g^2}{M^2} \alpha_0 \right)
                                \varphi^2
                        - \frac{g^2}{M^2} \alpha_4 \varphi \partial^4 \varphi
                        \\ \qquad
                        \mbox{} + \mbox{corrections at $\Mew^2$}
                \Big] ,
                \label{eq:effective-action}
        \end{eqnarray}
        and calculated only to tree level,
        where $\varphi$ represents any polarization of the vector boson of
        species $A$, and the corrections at $\Mew^2$ take the form of a tower
        of non-renormalizable terms suppressed by powers of $\Mew$.
        Note the unsuppressed non-renormalizable term
        of the form $\varphi \partial^4 \varphi$, which is symptomatic
    of the fact that our starting Lagrangian, Eq.~\eref{eq:action},
        did not describe a renormalizable quantum field theory.
        
        A good deal of information can be obtained from inspection
        of the effective action~\eref{eq:effective-action}.
        The relevant operators are the kinetic term
		$\varphi \partial^2 \varphi$
        and the mass term $\varphi^2$, which both receive corrections
        quadratic in the cutoff $\Lambda$.
        The mass is prevented from receiving corrections which scale
        faster than $\Lambda$
        because gauge invariance is restored when
        $M_A \rightarrow 0$, and in this limit the mass must not receive
        quantum corrections so that the Ward identity is preserved.
        Indeed, it follows from
        Eq.~\eref{eq:pi-expansion-a} that the $T$ parameter can be written
        \begin{equation}
                \alpha T = \frac{g^2}{M^2} \left(
                        \alpha_{0,Z} - \alpha_{0,W}
                \right) ,
        \end{equation}
        and therefore that all quadratic divergences cancel in this quantity.
        It is clear from Eq.~\eref{eq:effective-action}
        that this cancellation
        is a direct consequence of the restoration of gauge
        invariance in the limit $M_Z, M_W \rightarrow 0$.
        
        \subtitle{B. Low energy, massless vector bosons.}
        A similar procedure can be applied to find an effective Lagrangian
        for the photon self-energy in the low-energy limit.
        The vacuum polarization is obtained from Eq.~\eref{eq:self-energy}
        after the replacement $M_W \mapsto 0$, after which the
        expansions~\eref{eq:cutoff-corrections}--%
        \eref{eq:electroweak-corrections} continue to apply,
        with $\sigma^2$ substituted by the alternative combination
        $\tau^2$, which satisfies
        \begin{equation}
                \tau^2 \equiv 1 + x \frac{M_\chi^2}{\Lambda^2} .
        \end{equation}
        However, the roles of these non-renormalizable operators
        are subtly changed.
        Eq.~\eref{eq:cutoff-corrections} can still be interpreted
        as a tower of corrections at the cutoff
        (which we again caution will be accompanied by significant
        bridge contributions), but
        Eq.~\eref{eq:electroweak-corrections} now represents corrections
        at the scale of the dark energy mass, $|k^2| \sim M_\chi^2$.
        If we discard these corrections,
        it follows that the effective Lagrangian we obtain will be valid
        only in the limit $|k^2| \ll M_\chi^2$. Fortunately,
        for finite $M_\chi$ this is sufficient
        for the purpose of obtaining the oblique parameter $S$.
        
        In this limit, one finds
        \begin{equation}
                \Pi_{\gamma\gamma}(k^2) =
                        \frac{g^2}{M^2} \left[ \delta_2 k^2 + \delta_4 k^4 +
                        \Or\Big( \frac{k^2}{M_{\chi}^2} \Big) \right] ,
                \hfill ( |k^2| \ll M_\chi^2 ) \hspace{1cm}
        \end{equation}
        where the coefficients $\delta_2$ and $\delta_4$ satisfy
        \begin{eqnarray}
                \delta_2 \equiv \frac{\Lambda^2}{4}(1 - \Omega) +
                        \Or(M_\chi^2) ,
                \label{eq:delta-two} \\
                \delta_4 \equiv - \frac{1}{6}
                        + \Or\Big( \frac{M_\chi^2}{\Lambda^2} \Big) .
                \label{eq:delta-four}
        \end{eqnarray}
        Within its range of validity, this expansion can be interpreted
        in terms of the effective Lagrangian~\eref{eq:effective-action}.
        In particular, note that (as expected), no mass term is generated
        owing to gauge invariance.
        
        \subtitle{C. Energies near the resonance, massive vector bosons.}
        To obtain $S$, we require information about $\Pi_{AA}(k^2)$ in
        the region where it approaches the resonance at $k^2 = - M_A^2$.
        This can be studied by setting $k^2 = - M_A^2 + q^2$,
        and making an expansion in powers of $q^2/M_A^2$.
        When expanded in this way,
        it is less straightforward to interpret $\Pi(k^2)$
        as an effective Lagrangian.
        However, some of our understanding concerning
    	the meaning of each term can be carried over.
        
        Eqs.~\eref{eq:cutoff-corrections}--\eref{eq:electroweak-corrections},
        giving expansions in terms of non-renormalizable operators,
        continue to apply with the replacement $\sigma^2 \mapsto
        \hat{\sigma}^2$, where for each species $A$ of massive
        vector boson we have defined
        \begin{equation}
                \hat{\sigma}^2 \equiv 1 + (1-x)^2 \frac{M_A^2}{\Lambda^2}
                + x \frac{M_\chi^2}{\Lambda^2} .
        \end{equation}
        We find
        \begin{equation}
                \fl
                \Pi_{AA}(k^2) = \frac{g^2}{M^2} \left[
                        M_A^2 \hat{\alpha}_0 + \hat{\alpha}_2 q^2 + \hat{\alpha}_4 q^2
                        + \Or\Big( \frac{q^2}{\Mew^2}\Big)
                \right] ,
                \hfill
                (|k^2| \sim M_A^2)
                \hspace{1cm}
        \end{equation}
        where the coefficients $\hat{\alpha}_i$, for $i \in \{ 0, 2, 4 \}$,
        satisfy
        \begin{eqnarray}
                \fl\nonumber
                \hat{\alpha}_0 \equiv
                        \frac{\Lambda^2}{4} \left( \frac{\gamma^2}{2} -
                                \Omega(\epsilon-1) - 1 \right)
                \\ \mbox{}
                        + \frac{M_A^2}{36} \left(
                            3[\gamma(5\gamma - 6) + 4] \ln \frac{\Lambda^2}{M_A^2}
                                + 4[\gamma(4\gamma - 9) + 5]
                        + \Or \Big( \frac{M_\chi^2}{\Mew^2} \Big)
                \right) , \\
                \label{eq:alpha-hat-zero}
                \fl\nonumber
                \hat{\alpha}_2 \equiv
                        \frac{\Lambda^2}{4}(1 - \Omega)
                        + \frac{M_A^2}{72} \left(
                            3[ \gamma(11\gamma - 16) + 4 ] \ln \frac{\Lambda^2}{M_A^2}
                                + \gamma(67 \gamma - 128) + 59
                        + \Or \Big( \frac{M_\chi^2}{\Mew^2} \Big)       
                \right) , \\
                \label{eq:alpha-hat-two}
                \mbox{} \\
                \fl
                \hat{\alpha}_4 \equiv
                        \frac{1}{12} \ln \frac{\Lambda^2}{M_A^2} + \frac{11}{36}
                        + \Or \Big( \frac{M_\chi^2}{\Mew^2} \Big) .
                \label{eq:alpha-hat-four}
        \end{eqnarray}
        
        It is now possible to give expressions for the remaining
        oblique parameters $S$, $V$ and $W$ in terms of these effective
        quantities
        \begin{eqnarray}
                \frac{\alpha S}{4 \sw^2 \cw^2} =
                        \frac{g^2}{M^2} \left(
                                \alpha_{0,Z} - \hat{\alpha}_{0,Z} - \delta_2
                        \right) ,
                \label{eq:S-effective} \\
                \alpha V =
                        \frac{g^2}{M^2} \left(
                                \hat{\alpha}_{2,Z} + \hat{\alpha}_{0,Z} - \alpha_{0,Z}
                        \right) ,
                \label{eq:V-effective} \\
                \alpha W =
                        \frac{g^2}{M^2} \left(
                                \hat{\alpha}_{2,Z} + \hat{\alpha}_{0,W} - \alpha_{0,W}
                        \right) ,
                \label{eq:W-effective}
        \end{eqnarray}
        where we have dropped contributions from the non-renormalizable
        operator $\varphi \partial^4 \varphi$ since these never lead to
        quadratic divergences.
        It is now clear from inspection of
        Eqs.~\eref{eq:S-effective}--\eref{eq:W-effective}
        together with Eqs.~\eref{eq:alpha-zero}--\eref{eq:alpha-four},
        \eref{eq:delta-two}--\eref{eq:delta-four}
        and~\eref{eq:alpha-hat-zero}--\eref{eq:alpha-hat-four}
        that all quadratic divergences cancel in $S$, $T$, $U$
        $V$ and $W$.

        \section{Discussion}
        \label{sec:discuss}

        \subsection{When are quantum corrections screened?}
        \label{sec:constraints}

        This cancellation is not an accident, but is partly a consequence of
        gauge invariance
        and partly depends on the structure of gauge boson--lepton couplings
        within the Standard Model.

        The available phase space
 		which sets the size of the loop correction
		is determined by the couplings
        $\{ B, B_H \}$ and the mass of the boson, which is an infra-red effect.
        Most of the phase space volume will be concentrated near the
        ultra-violet region, in spherical shells
        of large Euclidean four-momentum. Coupling to these
        shells corresponds to a process
        where a propagating intermediate vector boson radiates into
        a hard chameleon and boson pair. From the point of view of
        this pair, the original vector boson behaves as if it were massless,
        and the effect of mass splittings between $W^\pm$, $Z$ and $\gamma$
        becomes irrelevant. Therefore, because gauge invariance requires that
        $W^\pm$, $Z$ and $\gamma$ couple to the dark energy in the same
        way at zero mass, we expect no difference in the manner
        in which any of these gauge bosons radiate into momentum shells
        at Euclidean four-momenta which are large compared with $M_Z$.

        Assuming our choice of input parameters,
        this is sufficient to screen all $\Or(1)$ effects
        in contact interactions of
        a single electroweak gauge boson with exactly
        two fermions---which is the only type of interaction which occurs in
        the electroweak sector, excluding interactions with the Higgs.
        The input parameters were chosen to be the $Z$ mass, $M_Z$,
        together with the fine structure constant,
        $\alpha$, and the Fermi constant, $G_F$, which measure the strength
        of the electromagnetic and charged-current interactions at zero
        momentum, respectively. Operationally, both $\alpha$ and $G_F$
        measure a combination of some dimensionless coupling constants
        and a propagator at zero momentum: for $\alpha$ this is the photon
        propagator, whereas $G_F$ measures the $W$ propagator.
        The oblique corrections can be of two kinds. Firstly, for processes
        involving a $Z$ particle, the strength of the neutral-current
        coupling is not measured by $G_F$ but can be obtained from it by
        a shift measured at zero momentum. This is the purpose of the $T$
        parameter.
        Secondly, a wavefunction renormalization of
        gauge boson lines may be necessary,
        which depends on properties of the
        propagator near Euclidean momentum of order $M_Z$.
        The difference between the wavefunction renormalization
        of the $Z$ and $W$ propagators evaluated
        at zero momentum and at momenta near $M_Z$ is measured
        by $V$ and $W$, respectively.
        Finally, $S$ compares the zero-momentum $Z$ and $\gamma$ propagators
        and therefore plays the same role for the photon as $T$ does
        for the $Z$, while $U$
        measures the difference between the $W^\pm$ and $Z$ propagators at
        zero momentum.

        All these shifts depend on a comparison of the phase space available
        to two different gauge bosons, or between the same gauge boson
        at different momenta. As we have seen, gauge invariance
        guarantees that the phase space available
        to all gauge bosons is the same at large Euclidean four-momentum,
        so differences can only arise from the interior shells of momentum
        space where the mass splitting between the electroweak gauge bosons
        can no longer be neglected. Differences in this region can not lead
        to $\Or(1)$ effects if the mass scales $\{ M$, $M_H \}$ associated with
        dark energy are much larger
        than the electroweak scale.
        It follows that large effects from radiative corrections are
        screened.
        However, this depends essentially on the fact that $\sqrt{\alpha}$
        and $\sqrt{G_F}$ include \emph{one} gauge boson line, and all other
        processes subsequently involve vertices which also include only
        a single ingoing or outgoing gauge boson. 
        
        The calculation of
    	the oblique corrections in the previous sections was done only
    	to one loop.  In principle loop corrections of any order could
    	contribute $\Or(1)$ corrections but we expect that the
    	screening of oblique scalar field corrections to the gauge
    	boson propagators occurs at all orders.

\begin{figure}
                \begin{center}
                        \includegraphics[scale=0.5]{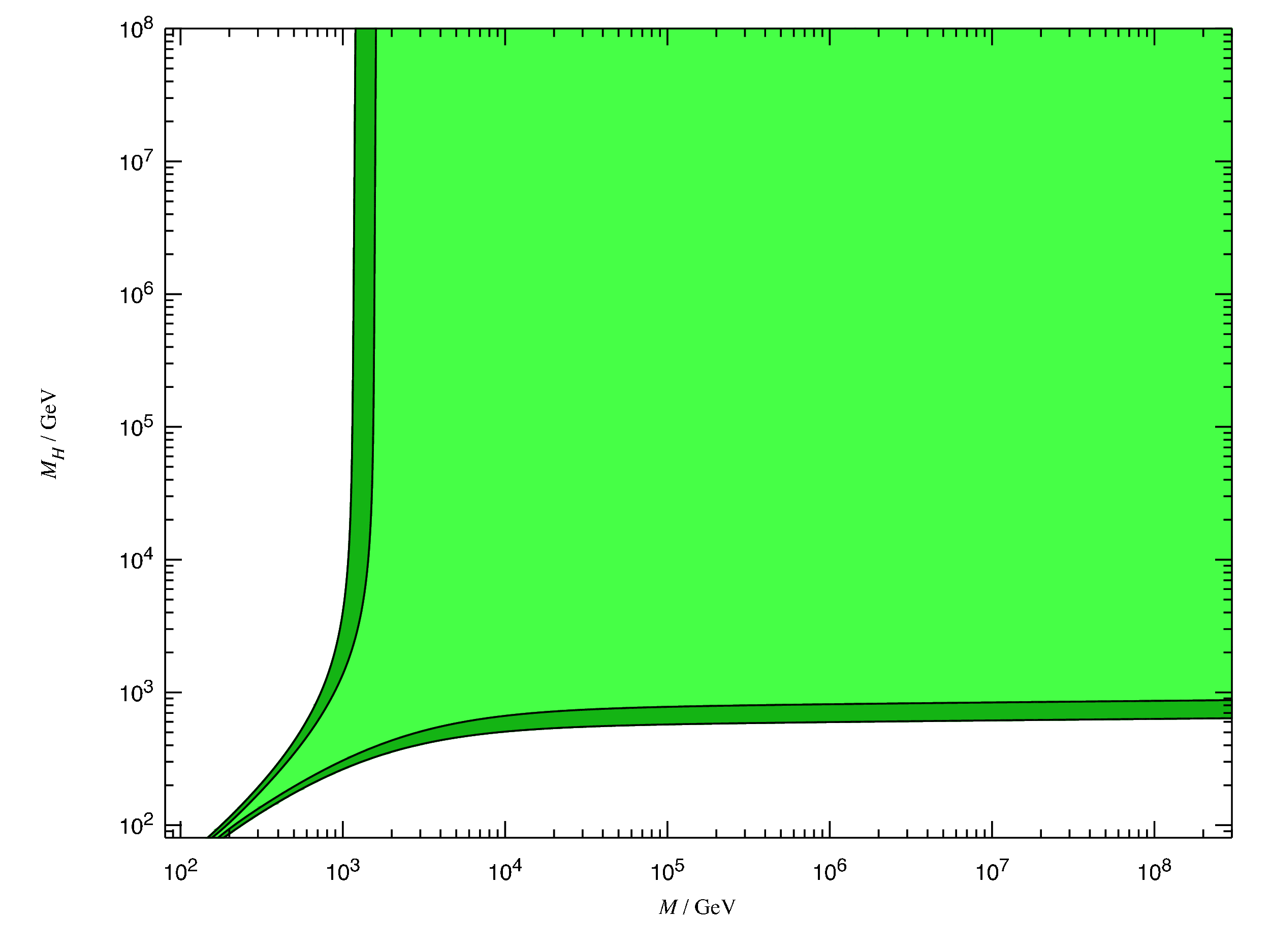}
                \end{center}
                \caption{\label{fig:constraints}Current collider constraints
                on the coupling scales $M$ and $M_H$, associated with dark
                energy interactions with the electroweak gauge and Higgs
                sectors, respectively. The interior light green region is
                compatible with current precision electroweak data at $1\sigma$,
                and extends indefinitely to large $M$ and $M_H$.
                Also shown is the $2\sigma$ region in darker green.}
        \end{figure}

        For a dark energy species which selects its mass via a
        chameleon mechanism
        we depict the current collider constraints on the
        mass scales $M$, $M_H$ in Fig.~\ref{fig:constraints}.
        \begin{figure}
                \begin{center}
                        \includegraphics[scale=0.5]{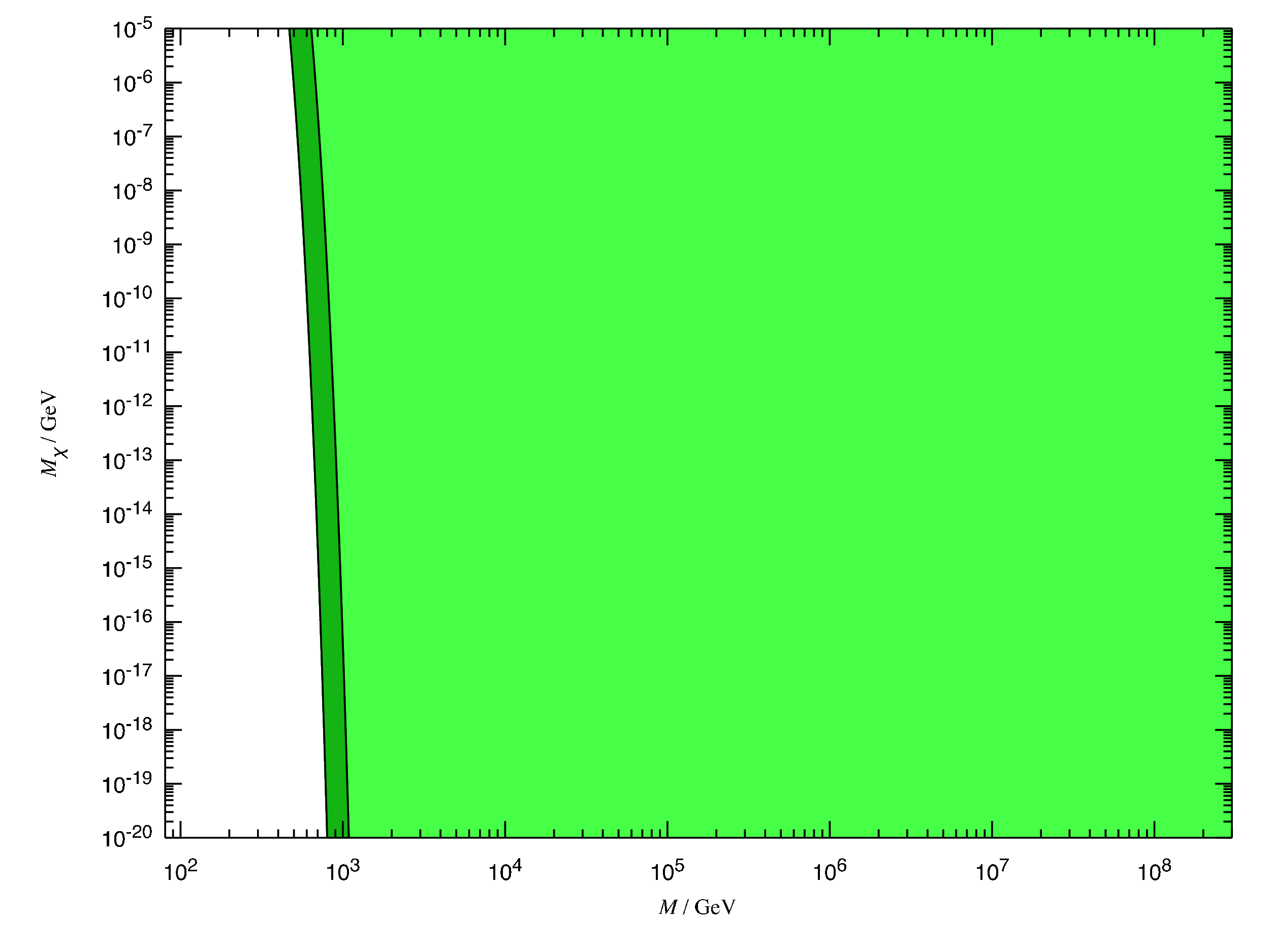}
                \end{center}
                \caption{\label{fig:masses}Current constraints on the
                mass, $M_\chi$, of a dark energy particle whose interaction
                with ordinary matter is characterized by a scale $M \gtrsim
                \Mew$. The interior light green region is compatible with
                present data at $1\sigma$, whereas the $2\sigma$ region is
                shown in darker green.
                Compare with Fig.~8 of Ref.~\cite{Andriamonje:2007ew}.}
        \end{figure}
        These can loosely be summarized as $M$, $M_H \gtrsim 1$ TeV,
        which is stronger than the constraint which follows from the
        decay width of the $Z$ into visible particles.
        Clearly, neither constraint is competitive with present bounds
        from optical or axion-search observations
        \cite{Chou:2008gr,Afanasev:2008jt}.
        In Fig.~\ref{fig:masses} we show the same constraints as a function
        of the dark energy mass, $M_\chi$, and its interaction scale
        $M$, without assuming that $M_\chi$ is determined by some
        chameleon-type mechanism.

        \subsection{Future prospects}
        \label{sec:future}

        Any future linear collider will
        measure electroweak precision observables
        with remarkable accuracy \cite{Weiglein:2007fp}, but if weak
        couplings imply it cannot
        produce dark energy particles directly
        then the most important discovery mode will come from
        sensitivity to radiative corrections at high energy. 
        For any electroweak processes sensitive to the diverging phase space
        of dark energy states at large Euclidean four-momentum, the
        discovery reach of the ILC would not be limited by the smallness
        of the coupling unless new physics operating at lower energies
        could quench the contribution of dark energy loops. An example of
        such new physics could be the appearance of a chameleon superpartner
        at some energy $\Msusy$, if $\Msusy \ll M$, where $M$ is the
        characteristic mass scale with which dark energy couples to the
        gauge sector.
        On the other hand, if dark energy radiative corrections \emph{are}
        screened, then contributions to electroweak precision observables
        fall with the mass scale of the coupling like
        $(\Mew/M)^2 \ln M^2/\Mew^2$.
        The most stringent constraint on $M$ currently
        derives from the polarization of light from astrophysical
        sources, which was studied in
        Refs.~\cite{Burrage:2008ii,Burrage:2009mj}
        and leads to the lower limit $M \gtrsim 10^9$ GeV. It is unlikely that
        such small corrections could ever be observed at the ILC.

\begin{figure}
                \vspace{5mm}
                \begin{center}
                        \begin{fmfgraph*}(120,60)
                                \fmfleft{l1,l2}
                                \fmfright{r1,r2,r3,r4}
                                \fmf{quark}{l2,vb}
                                \fmf{quark}{vb,r4}
                                \fmf{quark,left}{r3,r2}
                                \fmf{quark}{r1,vt1}
                                \fmf{boson,label=$W$}{vt1,vt2}
                                \fmf{quark}{vt2,l1}
                                \fmfforce{sw}{l2}
                                \fmfforce{nw}{l1}
                                \fmfforce{ne}{r1}
                                \fmfforce{se}{r4}
                                \fmfforce{(w,0.8h)}{r2}
                                \fmfforce{(w,0.2h)}{r3}
                                \fmffreeze
                                \fmf{quark,left=0.4}{vt1,vi}
                                \fmf{quark,left=0.4}{vi,vt2}
                                \fmf{gluon}{vi,vb}
                                \fmfv{label=$\bar{b}$}{l1}
                                \fmfv{label=$d$}{l2}
                                \fmfv{label=$d$}{r4}
                                \fmfv{label=$\bar{s}$}{r3}
                                \fmfv{label=$s$}{r2}
                                \fmfv{label=$\bar{s}$}{r1}
                                \fmfiv{label=$\}\hspace{2mm}\phi$}
                                          {(vloc(__r1) + vloc(__r2))/2 + (0.1w,0)}
                                \fmfiv{label=$\}\hspace{2mm}K^0$}
                                      {(vloc(__r4) + vloc(__r3))/2 + (0.1w,0)}
                                \fmfiv{label=$B^0$}
                                      {(vloc(__l1) + vloc(__l2))/2}
                        \end{fmfgraph*}
                \end{center}
                \caption{\label{fig:penguin}Penguin diagram contribution to the
                decay $B^0 \rightarrow K^0 \phi$.}
        \end{figure}
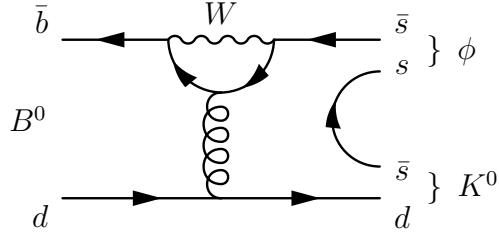

        Since detection of electroweakly interacting dark energy at
        $e^+ e^-$ colliders will be challenging,
        it is natural to consider what can be achieved at hadron--hadron
        colliders such as the LHC or the \emph{Tevatron}.
        Although $W^\pm$s and $Z$s are produced by such colliders,
        the problem of backgrounds and the difficulty of
        kinematical reconstruction of the final state at a hadron collider
        mean that constraints from pure electroweak processes are
        likely to be
        inferior to those from a future linear collider. However, hadron
        colliders are sensitive to other channels in which new physics can
        appear. One particularly interesting window on new physics may be
        provided by rare decays of $B$ mesons, which are bound states of
        a bottom quark $b$ with some other quark $q$ in the combinations
        $b \bar{q}$ or $\bar{b} q$. Such mesons can decay via flavour-changing
        neutral currents which are forbidden at tree-level, but give rise
        to decays such as $B^0 \rightarrow K^0 \phi$ (where $B^0$
        is the neutral $B$ meson composed of an anti-bottom/down pair
        $\bar{b}d$) when loop diagrams
        such as the so-called ``penguin'' of Fig.~\ref{fig:penguin}
        are included.
        Rare processes of this type
        give a comparatively clean signal of new physics.
        Unfortunately, it does not appear likely that conformally coupled
        dark energy of the kind studied in this paper could manifest
        itself in this way.
        At large Euclidean four-momentum, where the internal
        $W^\pm$ line in Fig.~\ref{fig:penguin} could be expected to
        receive sizeable dark energy modifications,
        the quarks to which it couples are effectively massless
        and the loop is flavour-independent.
        When summed over all quarks which can circulate in the loop,
        the unitarity of the Cabbibo--Kobayashi--Maskawa matrix
        implies that this dominant flavour-independent contribution
        suffers an exact cancellation: this is the
        so-called Glashow--Iliopoulos--Maiani mechanism. We can estimate that
        this mechanism allows
        any dark energy contribution, coupling at a scale $M$,
        to contribute at most at relative order $m_t^2/M^2$, where $m_t \approx
        175 \; \mbox{GeV}$ is the top mass.

        Is there any way to avoid the screening of large radiative corrections? 
        This can only be done if at least one
        coupling constant measured in a low-energy
        interaction can appear in a different context in some other process.
        Remarkably, the Standard Model does allow for this possibility.
        If we assume a minimal Higgs sector,
        there are three- and four-body interactions of the massive electroweak
        gauge bosons with physical Higgs quanta which are described by the
        action
        \begin{equation}
                \fl
                S = - \int \d^4 x \;
                        \left( 2^{1/4} \sqrt{G_F} H + G_F \frac{H^2}{\sqrt{2}} \right)
                        \left(
                                2 M_W^2 W^{+ a}W^-_a + M_Z^2 Z^a Z_a
                        \right) ,
        \end{equation}
        where $H$ is the physical Higgs field. This must itself be subject
        to oblique corrections which only involve the coupling $B_H$.
        There is no reason to expect that the shifts necessary to bring
        $G_F$ and the $H$ and gauge boson lines to finite momenta
        will be independent of the ultra-violet region of momentum space.
        However, these large effects are undetectable until the
        interaction of the Higgs with at least one of the massive gauge bosons
        becomes accessible to experiment. Even when this is possible, the
        details will depend sensitively on the mechanism of
        electroweak symmetry breaking chosen by Nature. For this reason
        we defer investigation of such processes,
        although we note that in the case of a minimal $SU(2)$
        doublet it is possible to verify that
        one could perhaps expect an $\Or(1)$ modification of the
        Higgs production rate via weak boson fusion at the LHC.

        The insensitivity of electroweak
        collider experiments to weakly coupled dark energy
        is frustrating given the inability
        of cosmological observations to place bounds on this region
        of parameter space. Although the search for astrophysical
        constraints has been fruitful \cite{Burrage:2007ew,Burrage:2008ii},
        it is difficult to imagine any astrophysical processes which would
        be sensitive to energy densities
        of order $(10^{12} \, \mbox{GeV})^4$ or above which were
        attained only during the very early universe. For example,
        one might have imagined that small perturbations imprinted in
        the dark energy scalar during an epoch of primordial inflation would
        lead to interesting effects in the late universe.
        Unfortunately, it appears that
        dark energy scalars of chameleon-type generically
        roll rapidly to their potential minima during the first few e-folds
        of inflation, where they remain for the duration of the accelerating
        era \cite{Brax:2004qh}. In the minimum, the dark energy field is
        heavy and is not excited by the inflationary expansion.
        For this reason,
        it does not function as an isocurvature field and cannot source
        evolution of the curvature perturbation, which might have led to
        interesting constraints from the spectral index or
        non-gaussianity.
        Moreover, the curvature perturbation is screened from possible
        non-perturbative effects because the chameleon vacuum expectation
        value is fixed \cite{Seery:2008ms}.
        On the other hand, if the
        chameleon vev shifted appreciably after inflation,
        it could potentially amplify the steep blue spectrum of perturbations
        imprinted on the $U(1)$ hypercharge field.
        If this amplification were too dramatic, it would lead to an
        unacceptable collapse of hypercharge fluctuations into primordial
        black holes at the end of inflation, in conflict with observation
        \cite{Carr:1993aq}. However, in practice the chameleon vev
        does not change sufficiently for this to provide an interesting
        constraint.

        \section{Conclusions}
        \label{sec:conclusions}

        In this paper, we have studied the prospects for collider
        physics to detect a scalar
        dark energy species which couples conformally to Standard Model states
        which are charged
        under the electroweak gauge group $SU(2) \times U(1)$.
        This is particularly interesting for proposals incorporating
        a chameleon-type mechanism, in which the dark energy field may
        evade stringent bounds on the presence of light scalar bosons
        by dynamically adjusting its mass to be large in regions of
        high average density.
        Any such theory of dark energy
        must certainly couple to the Standard Model, although
        it is not mandatory that the dark energy scalar couples to
        electroweak states. However, if such couplings \emph{are}
        present, then in view of the theoretical and experimental cleanliness
        of electroweak physics in comparison with hadron processes,
        one might expect
        that they would provide the most promising means of detection.
        
        In the minimal Standard Model with Higgs sector consisting of
        only a single $SU(2)$ doublet, this expectation is wrong.
        Although coupling to a dark energy scalar in principle allows for
        fractional shifts of $\Or(1)$ in precision electroweak observables,
        we have shown that in practice such large corrections are
        ``screened,'' in direct analogy with the screening theorem
        which prevents similar corrections from heavy Higgs states.
        Screening occurs because a majority of the dark energy
        corrections are absorbed in the input parameters
        $\{ \alpha, G_F, M_Z \}$, with only small splittings between
        the remaining observables which arise from the infra-red region
        of momentum space.
        The structure of the Standard Model also plays an important role,
        since all relevant vertices involve precisely two fermions and
        a single gauge boson. On the other hand,
        if it is possible to observe
        Higgs processes at the LHC then we would expect $\Or(1)$
        corrections to the Higgs production cross-section via
        weak-boson fusion.
        
        Although we have carried our explicit calculations only to one
        loop, we expect that screening of oblique corrections persists
        to all orders, any of which could contribute $\Or(1)$ effects as
        a matter of principle.
        This is important in establishing the consistency of
        dark energy theories
        with existing collider
        experiments, but also implies that the dark
		energy discovery potential of
        future $e^+ e^-$ colliders such as the proposed
        \emph{International Linear Collider} may be comparatively limited,
        unless the Higgs can be detected.
        
        One might also attempt to probe dark energy couplings via hadron
        processes, for which a promising observable might be the so-called
        flavour-changing neutral current which mediate rare decays of
        $B$ mesons. Unfortunately, for such reactions the unitarity of the
        Cabbibo--Kobayashi--Maskawa matrix plays a role similar to that
        of gauge invariance in quenching the contribution from shells of
        phase space at large Euclidean four-momentum, where a significant
        effect could be expected.
        Other non-electroweak effects such as bremsstrahlung also offer
        an apparently limited discovery potential.

        \ack
        
        It is a pleasure to thank Ben Allanach,
        Jeff Murugan and Malcolm Perry for helpful discussions.
        CB is supported by the German Science Foundation (DFG) under the
		Collaborative Research Centre (SFB) 676.
        ACD and DS are supported by STFC. AW is supported by
        the Cambridge Centre for Theoretical Cosmology.

        \appendix
        
        \section{Dark energy corrections to the Z width}
        \label{sec:z-calculation}
        
        In this Appendix we discuss the possibility of
        enhancements to the observed decay
        width of the neutral $Z$ boson. Such enhancements arise
        via the process $Z \rightarrow \chi Z^\ast \rightarrow \chi
        \psi \bar{\psi}$
        through which a propagating $Z$ emits a dark energy particle
        $\chi$ and passes off-shell. Eventually the off-shell $Z$ decays
        to a fermion--antifermion pair $Z^\ast \rightarrow \psi \bar{\psi}$,
		but
        if the weakly-interacting $\chi$ escapes the detector unseen
        then this reaction is indistinguishable from the direct decay
        $Z \rightarrow \psi \bar{\psi}$.
        
        \subsection{Direct decay}
        
        Let us first
		recapitulate the textbook calculation of direct decay
		\cite{Cheng:1985bj}.
		This will allow us
        to express the enhancement from dark energy emission as a fraction
        of the pure Standard Model rate. We suppose that a $Z$ particle decays
        into a fermion species whose quanta are created and destroyed
        by operators associated with the Dirac fields $\psi$ and $\bar{\psi}$,
        according to an interaction Lagrangian of the form
        \begin{equation}
                \int \d^4 x \;
                \bar{\psi} \gamma^a Z_a (g_L L + g_R R) \psi
        \end{equation}
        where the $\gamma^a$ are matrices obeying the Dirac algebra
        $\{ \gamma^a, \gamma^b \} = 2 \eta^{ab}$,
        $g_L$ and $g_R$ are arbitrary
        left- (respectively, right-) handed couplings,
        and $L$ and $R$ are projections onto the left- (respectively, right-)
        chirality halves of a spinor in Dirac's representation,
        \begin{equation}
                L \equiv \frac{1 + \gamma_5}{2}
                \quad \mbox{and} \quad
                R \equiv \frac{1 - \gamma_5}{2} .
        \end{equation}
        We use $\gamma_5 \equiv - \im \gamma^0 \gamma^1 \gamma^2 \gamma^3$,
        which has unit square $\gamma_5^2 = 1$ and commutes with all other
        $\gamma$-matrices. The projection operators $L$ and $R$
        obey $L^2 = L$ and $R^2 = R$, together with the orthogonality
        relation $LR = RL = 0$. We will also use the parity transformation
        operator $\beta = \im \gamma^0$, obeying $\beta^2 = 1$,
        in terms of which
        $\beta (\gamma^a)^\dag \beta = - \gamma^a$
        and
        $\beta \gamma_5 \beta = - \gamma_5$.
        
        Unpolarized decay proceeds according to the diagram of
        Fig.~\ref{fig:z-decay}\figlabel{a}.
        To obtain the overall rate, one
        averages over the three polarizations of a massive spin-1 particle
        and sums over the two spin states of each outgoing fermion.
        The differential decay rate
        per unit volume of phase space, $\d v$,
        available to the final state $\psi \bar{\psi}$
        pair corresponds to
        \begin{equation}
                \frac{\d \Gamma}{\d v} = 2 \pi
				\delta(\sum k)
                \sum_{\substack{\mathrm{outgoing} \cr \mathrm{spins}}}
                \frac{1}{3}
                \sum_{\substack{\mathrm{ingoing} \cr \mathrm{polarizations}}}
                | M_{\psi \bar{\psi}} |^2
                \label{eq:Z-decay-rate}
        \end{equation}
		where $\sum k$ schematically denotes the sum of all ingoing momenta
		minus all outgoing momenta.
        The Feynman amplitude $M_{\psi \bar{\psi}}$
        depends on the polarization of the initial $Z$, labelled
        $s$, and its 3-momentum $\vect{p}$,
        together with
        the spins of the final-state fermions, labelled
        $\sigma_{1,2}$, and their 3-momenta $\vect{t}_{1,2}$.
        It is defined by
        \begin{equation}
                \fl
                [M_{\psi \bar{\psi}}]^{s,\sigma_1 \sigma_2}
				\equiv
                - \frac{1}{(2\pi)^{3/2}}
                \left[ \bar{u}^{\sigma_1}(\vect{t}_1)
                        \gamma^a (g_L L + g_R R) v^{\sigma_2}(\vect{t}_2)
                \right]
                \frac{e_a^s(\vect{p})}{\sqrt{2 E_Z}} .
        \end{equation}
        After performing the spin and polarization sums,
 		yielding a trace over
        Dirac indices, this corresponds to a differential decay rate which can
        be written
        \begin{eqnarray}
                \fl\nonumber
                \d \Gamma = (2\pi)^4 \delta(p - t_1 - t_2)
                \frac{\d^3 t_1}{(2\pi)^3 2 E_1}
                \frac{\d^3 t_2}{(2\pi)^3 2 E_2}
                \frac{1}{6 E_Z}
                \\
                \mbox{} \times
                \left\{ 12 g_L g_R M_\psi^2 + (g_L^2 + g_R^2)
                \left( 4 \frac{(p \cdot t_1)(p \cdot t_2)}{M_Z^2} -
                        2 t_1 \cdot t_2 \right)
                \right\}
        \end{eqnarray}
        where $t_{1,2} = (E_{1,2},\vect{t}_{1,2})$
        and $p = (E_Z, \vect{p})$ are 4-momenta corresponding to the out-
        and in-going particles,
        respectively, and an infix dot `$\cdot$' denotes contraction
        in the Minkowski metric. All external particles are taken to be
		on-shell,
        with 4-momentum conservation enforced by
        $\delta(p - t_1 - t_2)$,
		and the outgoing fermions each have mass $M_\psi$.
        
        Conservation of 3-momentum is sufficient to determine one of the
        outgoing momenta. Moreover,
        performing the calculation in the $Z$ rest frame, symmetry requires
        that the outgoing fermions have equal energies $E_{1,2} = M_Z/2$.
        In consequence, we conclude that the total rate of emission
        into a solid angle $\d \Omega$ can be written
        \begin{equation}
                \fl
                \frac{\d \Gamma}{\d \Omega} =
                \frac{M_Z}{96 \pi^2}
                \sqrt{1 - 4 \frac{M_\psi^2}{M_Z^2}}
                \left\{
                        6 g_L g_R \frac{M_\psi^2}{M_Z^2}
                        + (g_L^2 + g_R^2)
						\left(1 - \frac{M_\psi^2}{M_Z^2} \right)
                \right\} .
                \label{eq:direct-decay}
        \end{equation}
        
        \subsection{Decay accompanied by dark energy emission}
        
        Now we return to the more complicated process where the decaying
        $Z$ is first pushed off-shell via emission of a single $\chi$ particle
        and subsequently decays into the observed fermion pair.
        This corresponds to the process of Fig.~\ref{fig:z-decay}\figlabel{b}.
        The $ZZ\chi$ interaction vertex is determined
		by~\eref{feynrule:w-cubic},
        modified as discussed below
        Eq.~\eref{eq:omega-def} to obtain the coupling to the $Z$ boson.
        
        As above
        we label the decaying $Z$ with momentum $\vect{p}$ and energy $E_Z$,
        and the outgoing fermions with momenta $\vect{t}_{1,2}$
        and energies $E_{1,2}$. The outgoing $\chi$ particle is taken to have
        momentum $\vect{q}$ and energy $E_\chi$.
        The total decay rate per unit of phase space available to the
        particles in the final state is given by a formula
        equivalent to Eq.~\eref{eq:Z-decay-rate}, with the
        Feynman amplitude
        $M_{\psi \bar{\psi}}$ replaced by
        a more complicated quantity
        $M_{\chi \psi \bar{\psi}}$ which satisfies
        \begin{eqnarray}
                \fl\nonumber
                [M_{\chi\psi\bar{\psi}}]^{s, \sigma_1 \sigma_2}
                \equiv
                - \frac{1}{(2\pi)^3} \frac{\bar{B}' M^{-1}}{r^2 + M_Z^2}
                \left[
                        \eta^{ab}(-p \cdot r - \gamma M_Z^2) + p^b r^a
                \right]
                \left( \eta_{bc} + \frac{r_b r_c}{M_Z^2} \right)
                \\ \mbox{} \times
                \left[
                        \bar{u}(\vect{t}_1)
                        \gamma^c G v(\vect{t}_2)
                \right]
                \frac{e_a^s(\vect{p})}{\sqrt{2 E_Z}}
                \frac{1}{\sqrt{2 E_\chi}} .
        \end{eqnarray}
        In order to avoid confusion with the parity inversion operator
        $\beta \equiv \im \gamma^0$ we have
        chosen the chameleon coupling
        scale as $M$, which elsewhere in this paper has been
        been synonymous with the
        coupling $\beta = M^{-1}$.
        The off-shell interior $Z$ carries 4-momentum
        $r = p - q$,
        and $G$ is the `coupling matrix,'
        \begin{equation}
                G_{\alpha \beta} \equiv
				\left[ g_L L + g_R R \right]_{\alpha \beta} ,
        \end{equation}
		where $\{\alpha, \beta, \ldots \}$ label spinor indices.
        Summing over final-state spins and averaging over all three
		initial-state
        polarizations, we find
        \begin{equation}
                \fl
                \d \Gamma = (2\pi)^4 \delta(p - q - t_1 - t_2)
                        \frac{\bar{B}^{\prime 2} M^{-2}}{(r^2 + M_Z^2)^2}
                        \frac{\d^3 q}{(2\pi)^3 2 E_\chi}
                        \frac{\d^3 t_1}{(2\pi)^3 2 E_1}
                        \frac{\d^3 t_2}{(2\pi)^3 2 E_2}
                        \frac{1}{6 E_Z}
                        \mathcal{M}'_{\chi \psi \bar{\psi}} ,
                \label{eq:accompanied-rate}
        \end{equation}
        where $\mathcal{M}'_{\chi \psi \bar{\psi}}$ satisfies
        \begin{equation}
                \fl
                \mathcal{M}'_{\chi \psi \bar{\psi}}
                \equiv
                {P^a}_c {P^d}_f 
                \left( \eta_{ad} + \frac{p_a p_d}{M_Z^2} \right)
                \tr \left\{
                        \gamma^c G (-1) (\im \slashed{t}_2 + M_\psi)
                        \beta (G)^\dag (\gamma^f)^\dag \beta
                        (-\im \slashed{t}_1 + M_\psi)
                        \right\}
                \label{eq:accompanied-decay} .
        \end{equation}
        We are adopting the usual Feynman convention in which
        $\slashed{Z} \equiv \gamma^a Z_a$ for any 4-vector $Z$;
        and ${P^a}_c$ is defined by
        \begin{equation}
                {P^a}_c \equiv
                        \left\{
                                \eta^{ab}(-p \cdot r - \gamma M_Z^2) + p^b r^a
                        \right\}
                        \left(
                                \eta_{bc} + \frac{r_b r_c}{M_Z^2}
                        \right) .
        \end{equation}
        The trace over Dirac indices can be evaluated by standard methods.
        It yields
        \begin{equation}
                \fl
                \mathrm{trace} =
                \eta^{cf} \left\{
                        4 g_L g_R M_\psi^2 - 2 (g_L^2 + g_R^2) t_1 \cdot t_2
                \right\} +
                2 (g_L^2 + g_R^2)(t_2^c t_1^f + t_2^f t_1^c) ,
                \label{eq:dirac-trace}
        \end{equation}
        plus a term antisymmetric under the exchange $c \leftrightarrow f$,
        which we omit because it disappears after insertion in
        Eq.~\eref{eq:accompanied-decay}.
        This trace depends only on the final $Z\psi\bar{\psi}$
        vertex and is common
        between the direct and accompanied decays. Nevertheless, it will not
        cancel in a ratio between the two, because it depends non-trivially
        on the Lorentz index structure by which it couples to the rest of the
        diagram. This structure receives significant modifications
        when the $Z$ decay is accompanied by dark energy emission.
        
        To proceed, we must contract Lorentz indices.
        We find
        \begin{eqnarray}
                \fl\nonumber
                \left( \eta_{ad} + \frac{p_a p_d}{M_Z^2} \right)
                {P^a}_c {P^d}_f =
                        p_c p_f (r^2 + \gamma^2 M_Z^2)
						+ \eta_{cf} ( p \cdot r +
                        \gamma M_Z^2)^2 \\
                \nonumber
                \mbox{} + r_c r_f
                \Bigg[ \left( \gamma - \frac{p \cdot r}{M_Z^2} \right)
                        \left( 1 + \frac{r^2}{M_Z^2} \right)
                        (p \cdot r + \gamma M_Z^2)  \\
                        \nonumber
                        \qquad\qquad \mbox{} +
                        \frac{1}{M_Z^2}(p \cdot r + \gamma M_Z^2)^2 +
                        \left( \frac{p \cdot r}{M_Z^2} \right)^2
                        (r^2 + \gamma^2 M)Z^2)
                \Bigg] \\
                \mbox{} + (r_c p_f + p_c r_f)
                        \left[ \frac{p \cdot r}{M_Z^2}(r^2 + \gamma^2 M_Z^2)
                                - \left( 1 + \frac{r^2}{M_Z^2} \right)
                                (p \cdot r + \gamma M_Z^2)
                        \right]
        \end{eqnarray}
        This can be contracted with Eq.~\eref{eq:dirac-trace} for the
        Dirac trace, yielding a final expression for
        $\mathcal{M}'_{\chi \psi \bar{\psi}}$.
        For convenience of expression,
        let us write $\mathcal{M}'_{\chi \psi \bar{\psi}} =
        \mathcal{A} + \mathcal{B}$. We find
        \begin{eqnarray}
                \fl\nonumber
                \frac{\mathcal{A}}
					{4 M_\psi^2 g_L g_R - 2(g_L^2 + g_R^2)t_1\cdot t_2}
                \equiv
                - M_Z^2(r^2 + \gamma^2 M_Z^2) + 4(p \cdot r + \gamma M_Z^2)^2
                \\ \nonumber
                \mbox{} +
                r^2 \Bigg[
                        \left( \gamma - \frac{p \cdot r}{M_Z^2} \right)
                        \left( 1 + \frac{r^2}{M_Z^2} \right)
                        ( p \cdot r + \gamma M_Z^2 )
                        \\ \nonumber
                        \qquad \qquad \mbox{}
                        + \frac{(p \cdot r + \gamma M_Z^2)^2}{M_Z^2}
                        + \left( \frac{p \cdot r}{M_Z^2} \right)^2
                                (r^2 + \gamma^2 M_Z^2)
                \Bigg]
                \\
                \mbox{} +
                2 (p \cdot r) \left[
                        \frac{p \cdot r}{M_Z^2}(r^2 + \gamma^2 M_Z^2) -
                                \left(1 + \frac{r^2}{M_Z^2} \right)
                                (p \cdot r + \gamma M_Z^2)
                \right] \label{eq:A-def}
        \end{eqnarray}
        and
        \begin{eqnarray}
                \fl\nonumber
                \frac{\mathcal{B}}{2(g_L^2 + g_R^2)} \equiv
                2 (p \cdot t_1)(p \cdot t_2)(r^2 + \gamma^2 M_Z^2) +
                2 (t_1 \cdot t_2)(p \cdot r + \gamma M_Z^2)^2
                \\ \nonumber
                \mbox{} + 2 (r \cdot t_1)(r \cdot t_2)
                \Bigg[ \left( \gamma - \frac{p \cdot r}{M_Z^2} \right)
                        \left( 1 + \frac{r^2}{M_Z^2} \right)
                        ( p \cdot r + \gamma M_Z^2 )
                        \\ \nonumber
                        \qquad \qquad \qquad \qquad \mbox{} +
                        \frac{(p \cdot r + \gamma M_Z^2)^2}{M_Z^2}
                        + \left( \frac{p \cdot r}{M_Z^2} \right)^2
                        (r^2 + \gamma^2 M_Z^2)
                \Bigg]
                \\ \nonumber
                \mbox{} + 2 [ (p \cdot t_1)(r \cdot t_2) + (p \cdot t_2)
                        (r \cdot t_1) ]
                        \\ \qquad \mbox{} \times
                        \left[
                                \frac{p \cdot r}{M_Z^2}(r^2 + \gamma^2 M_Z^2) -
                                \left( 1 + \frac{r^2}{M_Z^2} \right)
                                (p \cdot r + \gamma M_Z^2)
                        \right] \label{eq:B-def}
        \end{eqnarray}
        
        \subtitle{Kinematics.}
        As before, 3-momentum conservation is sufficient to determine the
        momentum of one outgoing particle, which we choose to be $\vect{t}_2$
        without loss of generality.
        Energy conservation determines one further scalar coordinate on phase
        space, which we choose to be $E_1$. The undetermined part of the
        3-body phase space is parametrized by the outgoing dark energy
        momentum $\vect{q}$ and a pair of polar and azimuthal angles
        $(\theta,\phi)$ which specify the orientation of $\vect{t}_1$ relative
        to $\vect{q}$.
        
        Let us obtain $E_1$ as a function of the unconstrained
        parameters. We work in the $Z$ rest frame. Energy conservation requires
        $E_\chi + E_1 + E_2 = M_Z$, and 3-momentum conservation fixes
        $\vect{t}_2 = - \vect{t}_1 - \vect{q}$.
        Therefore we conclude that $E_1$ must solve the implicit equation
        \begin{equation}
                E_2^2 = E_1^2 + E_\chi^2 - M_\chi^2 + 2
                \sqrt{E_\chi^2 - M_\chi^2}\sqrt{E_1^2 - M_\psi^2} \cos \theta .
                \label{eq:eone-implicit}
        \end{equation}
        To obtain an explicit equation, Eq.~\eref{eq:eone-implicit}
        can be squared and the resulting relation simplified. However,
        in doing so we introduce an extra solution for which
        Eq.~\eref{eq:eone-implicit} holds only after the exchange
        $\cos \theta \mapsto - \cos \theta$. The solution is spurious
        and should be eliminated. In practice we will find that the two
		possible
        solutions exchange roles at $\theta = \pi/2$, but that the matching
        is smooth.
        Following this procedure, the possible solutions must solve the
        quadratic equation
        \begin{eqnarray}
                \fl\nonumber
                E_1^2 \left[ 4 \cos^2 \theta (E_\chi^2 - M_\chi^2) -
                        4(M_Z - E_\chi)^2 \right]
                + 4 E_1 (M_Z - E_\chi)(M_Z^2 + M_\chi^2 - 2 M_Z E_\chi)
                \\
                = (M_Z^2 + M_\chi^2 - 2 M_Z E_\chi)^2
                + 4 M_\psi^2 \cos^2 \theta (E_\chi^2 - M_\chi^2) .
                \label{eq:eone-quadratic}
        \end{eqnarray}
        This can be achieved by substituting the correct solution of
        Eq.~\eref{eq:eone-quadratic} in
        Eqs.~\eref{eq:accompanied-rate}--\eref{eq:accompanied-decay}.
        However, one must also account for a Jacobian factor associated with
        transformation of the $\delta$-function enforcing
        energy conservation.
        To obtain this, define
        \begin{equation}
                \mathcal{E} \equiv
				E_Z - E_\chi - E_1 - E_2(E_\chi,E_1,\theta) ,
        \end{equation}
        in terms of which energy conservation requires $\delta(\mathcal{E})$.
        We can now use a change of variables to find
        \begin{equation}
                \delta(\mathcal{E}) =
                \frac{\delta[E_1 - E_1(E_\chi,\theta)]}
                {|\partial \mathcal{E}/\partial E_1|} ,
        \end{equation}
        where $E_1(E_\chi,\theta)$ is a solution of 
		Eq.~\eref{eq:eone-quadratic}.
        The Jacobian $\partial \mathcal{E}/\partial E_1$ can be determined
        using Eq.~\eref{eq:eone-implicit}, yielding
        \begin{equation}
                \left| \frac{\partial \mathcal{E}}{\partial E_1} \right|
                =
                \left|
                        1 + E_1 \frac{1 + (E_\chi^2 - M_\chi^2)^{1/2}
                                (E_1^2 - M_\psi^2)^{-1/2} \cos \theta}{E_2}
                \right| ,
        \end{equation}
        where $E_2$ is to be determined by Eq.~\eref{eq:eone-implicit}.
        In sum, the total unpolarized decay rate now satisfies
        \begin{equation}
                \frac{\d \Gamma}{\d \Omega_1}
                = \frac{\bar{B}^{\prime 2} M^{-2}}{192\pi^2 (2\pi)^3}
                \frac{\sqrt{E_1^2 - M_\psi^2}}{(r^2 + M_Z^2)^2}
                \frac{\d^3 q}{E_\chi E_2 M_Z}
                \left| \frac{\partial \mathcal{E}}{\partial E_1} \right|^{-1}
                \mathcal{M}'_{\chi \psi \bar{\psi}} ,
        \end{equation}
        where $\d \Omega_1$ is the element of solid angle
        associated with $\vect{t}_1$.
        
        To proceed it is convenient to introduce dimensionless
        small quantities $x$ and $y$, given in
        Eqs.~\eref{eq:x-def}--\eref{eq:y-def},
        which determine $M_\chi$ and $M_\psi$ in terms of $M_Z$,
        \begin{equation}
                M_\psi = \sqrt{x} M_Z
                \quad \mbox{and} \quad
                M_\chi = \sqrt{y} M_Z .
        \end{equation}
        Also, we can agree to measure energies in units of $M_Z$,
        introducing quantities $\hat{E}_\chi$ and $\hat{E}_{1,2}$
        which satisfy Eq.~\eref{eq:ehat-def}.
        Likewise, vectors such as $\vect{r}$ and $\vect{q}$ can be
        rescaled according to Eq.~\eref{eq:qhat-def},
        giving dimensionless vectors $\hat{\vect{r}}$ and
        $\hat{\vect{q}}$.
        In terms of these dimensionless quantities,
        Eq.~\eref{eq:direct-decay} giving the rate of direct decay reads
        \begin{equation}
                \frac{\d \Gamma(Z \rightarrow \psi \bar{\psi})}{\d \Omega}
                =
                \frac{M_Z}{96 \pi^2} \mathcal{M}_{\psi \bar{\psi}} ,
                \label{eq:dimensionless-direct-decay}
        \end{equation}
        where $\mathcal{M}_{\psi \bar{\psi}}$ satisfies
        Eq.~\eref{eq:direct-decay-matrix}
        with the fermion species $f$ taken to be $\psi$.
        The Jacobian
        $|\partial \mathcal{E}/\partial E_1|$
        satisfies Eq.~\eref{eq:jacobian-j},
        and we will denote it $J$ in what follows.
        Moreover, $\mathcal{M}'_{\chi \psi \bar{\psi}}$
        has mass dimension $[M^6]$, so we can introduce an unprimed
        quantity $\mathcal{M}_{\chi \psi \bar{\psi}}$ which
        depends only on $x$, $y$, the hatted vectors
        and other dimensionless quantities, and is defined by
        \begin{equation}
                \mathcal{M}'_{\chi \psi \bar{\psi}} \equiv
                M_Z^6 \mathcal{M}_{\chi \psi \bar{\psi}} .
        \end{equation}
        After these replacements, the rate of unpolarized decay accompanied
        by dark energy emission satisfies
        \begin{equation}
                \fl
                \frac{\d \Gamma(Z \rightarrow \chi \psi \bar{\psi})}
                {\d \Omega_1}
                =
                \frac{\bar{B}^{\prime 2} M^{-2}}{192 \pi^2 (2\pi)^3}
                M_Z^3
                \; \d \hat{E}_\chi \, \d \Omega_{\chi} \;
                \frac{\sqrt{\hat{E}_1^2 - x^2} \sqrt{\hat{E}_\chi^2 - y^2}}
                        {J (1 - \hat{E}_\chi - \hat{E}_1)(1+\hat{r}^2)^2}
                \mathcal{M}_{\chi \psi \bar{\psi}} .
                \label{eq:dimensionless-accompanied-decay}
        \end{equation}
        
        Taking the ratio of Eqs.~\eref{eq:dimensionless-accompanied-decay}
        and~\eref{eq:dimensionless-direct-decay}
        we finally obtain our advertised relation, Eq.~\eref{eq:rate-ratio},
        which describes the enhancement due to dark energy emission as a
        fraction of the bare Standard Model rate.
        As in Eq.~\eref{eq:total-ratio}, it is convenient to aggregate
        that part of the enhancement in
        Eq.~\eref{eq:dimensionless-accompanied-decay} which is independent of
        the chameleon coupling $M$ into a dimensionless integral
        $I_{\chi \psi \bar{\psi}}$. This will depend on the mass and
        couplings of the fermion species $\psi$, together with the mass of the
        dark energy scalar $\chi$. In particular,
        \begin{equation}
                \fl
                I_{\chi \psi \bar{\psi}}(M_\psi, M_\chi, g_L, g_R) \equiv
                \int \d \hat{E}_\chi \, \d \Omega_\chi \;
                \frac{\sqrt{\hat{E}_1^2 - x^2} \sqrt{\hat{E}_\chi^2 - y^2}}
                        {J (1 - \hat{E}_\chi - \hat{E}_1)(1+\hat{r}^2)^2}
                \frac{\mathcal{M}_{\chi \psi \bar{\psi}}}
                        {\mathcal{M}_{\psi \bar{\psi}}} .
                \label{eq:I-def}
        \end{equation}
        We give representative values for $I_{\chi \psi \bar{\psi}}$ in
        Table~\ref{table:Ivals}.
        \begin{table}
                \begin{center}
                        \tiny
                        \renewcommand{\arraystretch}{1.7}
                        \begin{tabular}{c|ccccc} \hline \hline
                                \textbf{\textsf{fermion species}} &
                                massless neutrino &
                                $10^{-3}$~eV neutrino &
                                $511$~keV electron &
                                5~GeV electron &
                                40~GeV electron \\ \hline
                                $I_{\chi \psi \bar{\psi}}$ & 
                                $0.22$ & $0.22$ & $0.22$ & $0.21$ & $0.007$ \\
                                \hline\hline 
                        \end{tabular}
                \end{center}
                \caption{\label{table:Ivals}Enhancement factors for $Z$
                decay accompanied by dark energy emission, to be interpreted in
                conjunction with Eqs.~\eref{eq:total-ratio}
                and~\eref{eq:I-def}. As the fermion mass increases, the phase
                space available to any decay products diminishes until
                it is forbidden altogether at the kinematic threshold
                $M_\psi = M_Z/2$.
                The enhancement for massless or light particles is
                very nearly independent of their identity.}
        \end{table}
        
        \section{Bridges and daisies: dark-energy corrections to
        all orders}
        \label{sec:bridges}
        
        Even in the effective field theory interpretation, where loops
        which are purely internal to the dark energy sector are ignored,
        Eq.~\eref{eq:action}---together with the assumption that all
        matter fields couple conformally---entails a great many possible
        corrections to Standard Model processes. In this Appendix, we
        argue that to an acceptable approximation
        a majority of these corrections are subdominant; in this approximation,
        the leading dark energy effect comes from the one-loop
        oblique correction. This assumption played an essential role
        in determining the dark energy corrections in \S\ref{sec:ewpo}.
        
        A useful example to keep in mind is the case of the graviton,
        which also couples conformally to matter (and indeed
        all Standard Model states) with a universal coupling function,
        $\sqrt{\det(\eta_{ab} + h_{ab})}$, where $\eta_{ab}$ is the background
        metric and $h_{ab}$ is the spin-2 graviton field. This non-linear
        coupling leads to a network of gravitational bridges, daisies
        and oblique corrections,
        quite analogous to Fig.~\ref{fig:dark-energy-processes},
        which also dress Standard Model processes. In the case of gravitons these
        have very little impact on reactions taking place at collider energies;
        in comparison,
        the structure of the dark energy interactions in
        Eq.~\eref{eq:action} allow a small number of diagrams to make an
        $\Or(1)$ contribution. Nevertheless, many similarities exist between
        graviton and dark energy phenomenology.
        
\begin{figure}
                \vspace{3mm}
                \hfill
                \begin{fmfgraph*}(175,60)
                        \fmfleft{l1,l2}
                        \fmfright{r1,r2}
                        \fmf{fermion}{l1,v1}
                        \fmf{fermion}{l2,v1}
                        \fmf{boson}{v1,v2}
                        \fmf{boson}{v2,v3}
                        \fmf{boson}{v3,v4}
                        \fmf{fermion}{v4,r1}
                        \fmf{fermion}{v4,r2}
                        \fmffreeze
                        \fmf{plain,left=0.5}{v1,v2}
                        \fmf{plain,left=1.0}{v1,v2}
                        \fmf{plain,left=1.5}{v1,v2}
                        \fmf{plain,right=0.75}{v2,v3}
                        \fmf{plain,right=1.50}{v2,v3}
                        \fmf{plain,left=0.375}{v3,v4}
                        \fmf{plain,left=0.750}{v3,v4}
                        \fmf{plain,left=1.125}{v3,v4}
                        \fmf{plain,left=1.500}{v3,v4}
                \end{fmfgraph*}
                \figlabel{a}
                \hfill
                \begin{fmfgraph*}(100,60)
                        \fmfleft{l1,l2}
                        \fmfright{r1,r2}
                        \fmf{fermion}{l1,v1}
                        \fmf{fermion}{l2,v1}
                        \fmf{boson}{v1,v2}
                        \fmf{fermion}{v2,r1}
                        \fmf{fermion}{v2,r2}
                        \fmffreeze
                        \fmf{plain,left=0.5,tag=1}{v1,v2}
                        \fmf{plain,left=1.5,tag=2}{v1,v2}
                        \fmfposition
                        \fmfipath{p[]}
                        \fmfiset{p1}{vpath1(__v1,__v2)}
                        \fmfiset{p2}{vpath2(__v1,__v2)}
                        \fmfi{boson}{point length(p1)/2 of p1 --
                                     point length(p2)/2 of p2}
                \end{fmfgraph*}
                \figlabel{b}
                \hfill
                \mbox{}
                \\[8mm]
                \mbox{}
                \hfill
                \begin{fmfgraph*}(150,60)
                        \fmfleft{l1,l2}
                        \fmfright{r1,r2}
                        \fmf{fermion}{l1,v1}
                        \fmf{fermion}{l2,v1}
                        \fmf{boson,tension=1.5}{v1,v2}
                        \fmf{boson,tension=1.5}{v2,v3}
                        \fmf{boson}{v3,v4}
                        \fmf{boson,tension=1.5}{v4,v5}
                        \fmf{boson,tension=1.5}{v5,v6}
                        \fmf{fermion}{v6,r1}
                        \fmf{fermion}{v6,r2}
                        \fmffreeze
                        \fmf{plain,left=0.50}{v2,v5}
                        \fmf{plain,left=0.75}{v2,v5}
                        \fmf{plain,left=1.00}{v2,v5}
                        \fmf{plain,right=0.75}{v3,v4}
                        \fmf{plain,right=1.25}{v3,v4}
                \end{fmfgraph*}
                \figlabel{c}
                \hfill
                \mbox{}
                \caption{\label{fig:complex-bridges}More complicated bridge-class
                diagrams. In addition to the simple bridge shown in
                Fig.~\ref{fig:dark-energy-processes}\figlabel{c},
                one can use high-order vertices between two gauge bosons and
                an arbitrary number of dark energy quanta to ``chain'' any
                number of bridges together, as shown in \figlabel{a}.
                In \figlabel{b}, the component lines of a particular bridge
                are themselves joined together by virtual quanta of other
                species. (These could include dark energy particles, because
                the loops formed by such ``joined bridges'' would not already be
                included in the effective dark energy sector.)
                In \figlabel{c}, bridge diagrams are nested within each other
                to form so-called rainbow diagrams.}
        \end{figure}
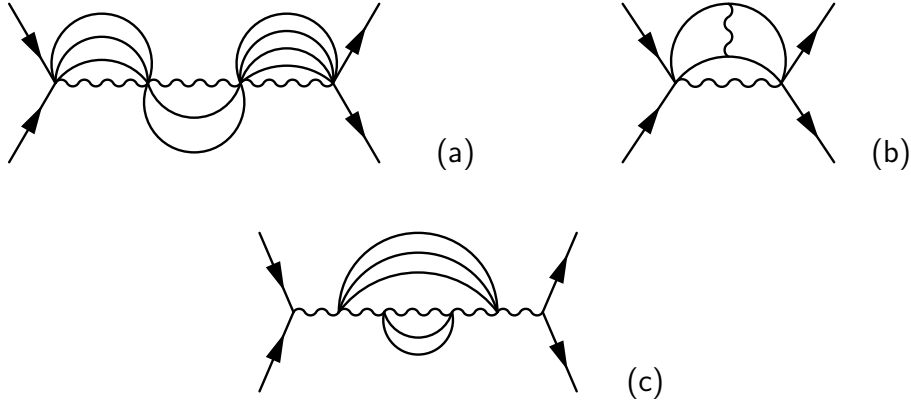

        Together with the simple daisy and bridge classes introduced
        in Fig.~\ref{fig:dark-energy-processes}, one can contemplate
        more complicated corrections.
        Bridges can be \emph{chained} together in arbitrary combinations,
        as shown in Fig.~\ref{fig:complex-bridges}\figlabel{a},
        or the component lines within a given bridge can themselves be
        joined together by other particles, as in
        Fig.~\ref{fig:complex-bridges}\figlabel{b}.
        Alternatively, bridges can be nested within each other to create
        \emph{rainbows}---see Fig.~\ref{fig:complex-bridges}\figlabel{c}.
        In principle, a hierarchy of resummations (somewhat similar to the
        Balitsky hierarchy in QCD) is necessary to accommodate
        all these types of activity.
        
        \subsection{Daisy diagrams}
        Let us first consider the effect of daisies which dress bare
        Standard Model vertices. These are always momentum-independent
        and merely constitute a renormalization of whichever coupling
        constant sets the strength of the interaction at the vertex.
        For this reason they are relatively straightforward to deal with,
        and in the simplest situation we shall be able to resum their
        effect to all orders.
        If the daisies vary between different species of fermion, then
        the result would be an apparent species-dependent Fermi constant,
        $G_F$. To prevent this occurring the fermion coupling function
        must be universal, which will be the case for conformal couplings.
        In what follows, we assume this to be the case.
        
        Consider Eq.~\eref{eq:action} and expand the coupling functions
        $B(x)$ and $B_H(x)$ according to
        \begin{eqnarray}
                B(\beta \chi) \equiv \sum_{n=0}^\infty \frac{1}{n!} \bar{B}_n
                \beta^n (\delta \chi)^n \\
                B_H(\beta \chi) \equiv \sum_{n=0}^\infty \frac{1}{n!} \bar{B}_{H,n}
                \beta_H^n (\delta \chi)^n ,
        \end{eqnarray}
        where $\chi = \bar{\chi} + \delta\chi$, given that $\bar{\chi}$
        is the expectation value of the dark energy scalar in the vacuum,
        and $\bar{B}_n$ ($\bar{B}_{H,n}$) are the Taylor coefficients of $B$
        ($B_H$) evaluated in this vacuum.
        We assume that $\bar{B}_0 = \bar{B}_{H,0} = 1$ and
        introduce a quantity $\gamma_n$, defined by
        \begin{equation}
                \gamma_n \equiv \frac{\bar{B}_{H,n} \beta_H^n}{\bar{B}_n \beta^n} .
        \end{equation}
        In terms of $\bar{B}_n$ and $\gamma_n$, the $n$th order interaction
        vertex takes the form
        \begin{eqnarray}
                \fl\nonumber
                S_n = \int \frac{\d^4 k_1}{(2\pi)^4}
                \frac{\d^4 k_2}{(2\pi)^4}
                \frac{\d^4 p_1}{(2\pi)^4}
                \cdots
                \frac{\d^4 p_n}{(2\pi)^4}
                (2\pi)^4 \delta(k_1 + k_2 + \sum_j p_j)
                \\
                \frac{\bar{B}_n \beta^n}{n!} W_a^+(k_1) W_b^-(k_2)
                \delta \chi(p_1) \cdots \delta \chi(p_n)
                [ \eta^{ab}(k_1 \cdot k_2 - \gamma_n M_W^2) - k_1^b k_2^a ] .
        \end{eqnarray}
        As an example,
        we will compute the simplest class of daisies which contribute to
        the interior of the $W^\pm$ propagator.
        The calculation of daisies which dress vertices with other
        species of fermion---or for the other gauge bosons---proceeds analogously.
        One finds that the vacuum polarization with momentum
        transfer $q$, which arises from the daisy with $n$
        petals, can be written
        \begin{equation}
                \fl
                \Pi_{ab}^n \supseteq - 
                        \bar{B}_{2n} \beta^{2n} \frac{(n-1)!!}{(2n)!}
                        \left\{ \int \frac{2\pi^2 \kappa^3 \; \d \kappa}{(2\pi)^4}
                        \frac{1}{\kappa^2 + M_\chi^2} \right\}^n
                        [ \eta_{ab}(q^2 + \gamma_n M_W^2) - q^a q^b ] ,
                \label{eq:daisy-all}
        \end{equation}
        where $(n-1)!! \equiv (n-1)(n-3) \cdots 1$ is the so-called
        ``double factorial.'' In the special case of an exponential coupling,
        where $\bar{B}_n = 1$ for all $n$, and assuming that
        $\gamma_n$ can be replaced by a constant $\gamma$, then
        it follows that all orders of daisies can be resummed to give
        \begin{eqnarray}
                \fl\nonumber
                \Pi_{ab} \supseteq -
                [ \eta_{ab}(q^2 + \gamma M_W^2) - q_a q_b ]
                \Bigg\{ - 1 +
                        F\Big(  \begin{array}{c}
                                        \cdot \\
                                        1/4, 3/4
                                \end{array} \Big| \frac{(\beta \Lambda)^4}{2^{15} \pi^4}
                                \left[1 - \frac{M_\chi^2}{\Lambda^2}
                                \ln \frac{\Lambda^2}{M_\chi^2} \right]^2
                        \Big)
                        \\ \nonumber \mbox{} +
                        \frac{(\beta \Lambda)^2}{32\pi^2}
                        \left( 1 - \frac{M_\chi^2}{\Lambda^2} \ln
                                \frac{\Lambda^2}{M_\chi^2} \right)
                        F\Big(  \begin{array}{c}
                                        1 \\ 3/4, 5/4, 3/2
                                \end{array} \Big| \frac{(\beta \Lambda)^4}{2^{15} \pi^4}
                                \left[1 - \frac{M_\chi^2}{\Lambda^2}
                                \ln \frac{\Lambda^2}{M_\chi^2} \right]^2
                        \Big)
                \Bigg\}
                \\
                \label{eq:daisy-resum}
        \end{eqnarray}
        where $F(a; b | z)$ is the generalized hypergeometric function.
        For $\Lambda \lesssim \beta^{-1}$ and $M_\chi \ll \Lambda$
        this resummation is dominated by its one-loop term.
        There will be extra terms in addition
        to Eq.~\eref{eq:daisy-resum} which arise from
        interference between daisy diagrams and bridge diagrams
        (to be discussed in the next section).
        Although these may change
        the details of some numerical coefficients, they will not
        alter the momentum-independent character of the corrections.
        
        Eqs.~\eref{eq:daisy-all}--\eref{eq:daisy-resum} exhibit the general
        features which will recur in all diagrams we consider in this
        Appendix. A diagram with $n$ dark energy lines can contribute
        at leading order in powers of $\beta \Lambda$, with this contribution
        coming from a region of phase space where all dark energy lines
        are carrying momenta of order $\Lambda$. This would seem to suggest
        that diagrams containing an arbitrary number of lines
        need to be accounted for in order to make reliable predictions.
        However, each line is also accompanied by a phase-space
        factor of $1/16\pi^2$ (plus the combinatorical factor
        $(n-1)!!/(2n)!$) which leads to suppression of high-loop terms,
        so that counting powers of $\beta \Lambda$ alone does not give
        a proper accounting of the relative magnitude of
        adjacent terms in the loop expansion. 
        
        \subsection{Bridge diagrams}
        Bridge diagrams are more complicated to handle.
        We proceed in two steps, first arguing that a similar phase-space
        suppression means that only the one-loop bridge need be considered,
        and not multi-loop or rainbow bridges.
        In a second step, we argue that \emph{chains}
        of bridges can be ignored because they make a contribution at
        leading order in powers of $\beta \Lambda$ which is precisely
        momentum-independent. A contribution of this type can be
        absorbed into coupling constants and becomes unobservable.
        The one-loop bridge is
        itself momentum-independent at leading order in powers of
        $\beta \Lambda$, so that the only contributions at this order
        which are not suppressed by a phase space factor of order
        $\sim 100$ are the oblique corrections considered in
        \S\ref{sec:ewpo}. It will transpire that we expect
        corrections to the purely oblique analysis of
        \S\S\ref{sec:ewpo}--\ref{sec:discuss} to occur at a relative order
        of roughly $1/8\pi^2 \approx 0.013$ or better.
        We believe this is an acceptable
        precision at which to predict what can be observed at present and
        future particle colliders.
        It is again important that there is a universal fermion coupling
        function, in order that the effective Fermi constant $G_F$ does not
        become species-dependent.
        
        First consider a multi-loop contribution to the vacuum polarization
        of any $SU(2)$ gauge boson. At momentum transfer $q$,
        a calculation similar to those presented in \S\ref{sec:polarization}
        establishes that this can be represented in the form
        \begin{eqnarray}
                \fl\nonumber
                \Pi_{ab}^n \supseteq \frac{\bar{B}_n^2 \beta^{2n}}{\bar{B}_1 n!}
                (-\im)^n \int \frac{\d^4 \ell}{(2\pi)^4}
                \frac{\d^4 r_1}{(2\pi)^4}
                \cdots
                \frac{\d^4 r_{n-1}}{(2\pi)^4}
                \\ \mbox{} \times
                \frac{P_{ab}(\ell,q)}{\ell^2 + M_W^2 - \im \epsilon}
                \frac{1}{r_1^2 + M_\chi^2 - \im \epsilon}
                \cdots
                \frac{1}{r_{n-1}^2 + M_\chi^2 - \im \epsilon}
                \frac{1}{R^2 + M_\chi^2 - \im \epsilon} ,
                \label{eq:multiloop-bridge}
        \end{eqnarray}
        where $R = q + \ell + \sum_{j = 1}^{n-1} r_j$,
        $\ell$ is the momentum carried by the exchanged gauge boson,
        and $P_{ab}(\ell,q)$ satisfies
        \begin{equation}
                \fl
                P_{ab} \equiv \eta_{ab}( \ell \cdot q - \gamma_n M_W^2 )^2
                - (\ell_a q_b + \ell_b q_a)(\ell \cdot q - \gamma_n M_W^2)
                + \ell_a \ell_b (q^2 + \gamma_n^2 M_W^2) .
        \end{equation}
        Eq.~\eref{eq:multiloop-bridge}
        is suppressed by $2n$ powers of the coupling
        $\beta$, but each scalar integral can contribute a power
        of $\Lambda^2$. Since $P_{ab} \sim \ell^2$, the $\ell$
        integration can contribute terms of order
        $\Lambda^4$. This would seem to imply that terms of order
        $\Lambda^2 (\beta \Lambda)^{2n}$ could be present in the answer,
        but the correct conclusion depends on the relative magnitude of
        $R^2$. Unlike the daisy diagrams or one-loop bridges, there can be
        some regions of phase space where
        $\ell \sim r_j \sim \Lambda$ and
        $|\ell + \sum_{j = 1}^{n-1} r_j| \sim 0$, so that
        $R \sim q$. However, if the region of phase space in which this
        finely-tuned cancellation occurs shrinks with increasing cutoff
        faster than $\Lambda^2$ then we can estimate the leading contribution
        by setting $R \sim \Lambda$ in Eq.~\eref{eq:multiloop-bridge}.
        In practice, the enhanced region of phase space is negligibly small.
        
        \paragraph{The $n$-loop bridge.}
        To estimate the contribution of the $n$-loop bridge, we set
        $R \sim n \Lambda$ and replace each factor such as
        $\d^4 r_j/(2\pi)^4$ by $\Lambda^2/8\pi^2$.
        This choice for $R$ is tantamount to assuming that the
        $r_j$ and $\ell$ are randomly oriented, so that their cross terms
        approximately average to zero. This is likely to be a good
        approximation for large $n$ but may fail for $n \sim \Or(1)$,
        so we will demonstrate explicitly that this procedure gives the correct
        answer for the 2-loop bridge.
        We consider this to be reasonable evidence that our estimate
        is reliable for all $n$.
        Proceeding in this way, it follows that the $n$-loop
        bridge makes a contribution to the vacuum polarization
        which is roughly equal to
        \begin{equation}
                \Pi_{ab}^n \supseteq \frac{1}{4} \eta_{ab}
                        (2 q^2 + \gamma_n^2 M_W^2)
                        \frac{\bar{B}_n^2 (\beta \Lambda)^{2n}}{\bar{B}_1 n \cdot n!}
                        \left( \frac{1}{8\pi^2} \right)^{n} .
        \end{equation}
        For the special case of an exponential coupling---for which
        $\bar{B}_n = 1$ for all $n$---this can be
        resummed to give
        \begin{equation}
                \Pi_{ab} \supseteq
                        \frac{1}{4} \eta_{ab} ( 2 q^2 + \gamma^2 M_W^2 )
                        \left\{ - \gamma_E + \Ei(\frac{\beta^2 \Lambda^2}{8\pi^2} )
                                - \ln \frac{\beta^2 \Lambda^2}{8\pi^2} \right\} ,
                \label{eq:pol-estimate}
        \end{equation}
        where $\gamma_E \approx 0.577$ is the Euler--Mascheroni constant
        and $\Ei(z)$ is the exponential integral. As in the case of
        the daisy diagrams this is dominated by its one-loop contribution,
        with the contribution of higher loops being suppressed by
        the phase-space factor $1/8\pi^2$.
        
        \paragraph{Two-loop bridge.} In the special case of the two-loop
        bridge, Eq.~\eref{eq:multiloop-bridge} can be evaluated by the
        method of Feynman parameters, as in \S\ref{sec:polarization},
        with the result that
        \begin{eqnarray}
                \fl\nonumber
                \Pi_{ab}^{(2)} = \frac{\bar{B}_2^2 \beta^4}{\bar{B}}
                \left( \frac{1}{8\pi^2} \right)^2
                \int_0^{1} \d x \, \d y \, \d z \; \delta(1-x-y-z)
                \\ \nonumber \mbox{} \times
                \int_0^\Lambda \ell^3 \, \d \ell
                \int_0^\Lambda r^3 \, \d r \;
                \frac{\frac{1}{4} \ell^2 (2q^2 + \gamma_2^2 M_W^2) +
                  (X q^2 + \gamma_2 M_W^2)^2}
                     {\Gamma^3} ,
        \end{eqnarray}
        where $\Gamma$ is defined by
        \begin{equation}
                \Gamma \equiv (1-x) r^2 + W \ell^2 + X Y q^2 + x M_W^2 +
                (1-x) M_\chi^2 ,
                \label{eq:Gamma-def}
        \end{equation}
        and the three quantities $W$, $X$ and $Y$ satisfy
        \begin{eqnarray}
                W \equiv \frac{(1-x)(1-y)-z^2}{1-x} , \\
                X \equiv \frac{z(1-z)}{(1-x)(1-y)-z^2} , \\
                Y \equiv \frac{(1-x)(1-y)-z^2 + z(1-z)}{1-x} .
        \end{eqnarray}
        Performing the $\ell$ and $r$ integrals and keeping only the leading
        term in powers of $\beta \Lambda$, this vacuum polarization can be
        simplified to read
        \begin{eqnarray}
                \fl\nonumber
                \Pi_{ab}^{(2)} \simeq
                        -\frac{\bar{B}_2^2 (\beta \Lambda)^4}{\bar{B}}
                        (2 q^2 + \gamma_2^2 M_W^2)
                        \frac{1}{32}
                        \left( \frac{1}{8\pi^2} \right)^2
                        \int_0^1 \d x \, \d y \, \d x \; \delta(1 - x - y - z)
                        \\ \mbox{} \times
                        \frac{1}{W^3(W+1-x)}
                        \left[
                                W + (W+1-x) \ln \frac{1-x}{W+1-x}
                        \right] ,
                \label{eq:two-loop-pol}
        \end{eqnarray}
        where by ``$\simeq$'' we mean that this relationship is true up to
        terms of order $\beta^2 (\beta \Lambda)^2$ which we have neglected.
        The possibility of enhanced regions of phase space where
        the loop momenta approximately cancel to leave an anomalously
        small propagator $\sim 1/q^2$ (rather than
        $\sim 1/\Lambda^2$) has been replaced by
        the potential for large contributions from the
        Feynman parameter integrals.
        Indeed, inspection of Eq.~\eref{eq:Gamma-def} shows when considering
        only the
        \emph{leading} term in powers of $\Lambda$ we might find
        a divergence roughly like $\int \d x / (1-x)^3$.
        This would be finite when terms of all orders in $\Lambda$
        were included, but would manifest as an apparent divergence in the
        truncated series. If such a divergence appears,
        it should be regulated at a scale roughly of order
        $M_\chi^2/\Lambda^2$ where other terms in the perturbation theory
        become important, allowing enhanced phase space regions to appear.
        However, when these integrals are
        treated sufficiently carefully we find that no divergences occur and
        therefore that no enhanced regions of phase space exist.
        The integral can be evaluated by an adaptive Monte Carlo technique,
        and we find its numerical value to be roughly $\approx 0.025$.
        We conclude that Eq.~\eref{eq:two-loop-pol} is as small in
        magnitude, or slightly smaller, than our estimate
        Eq.~\eref{eq:pol-estimate}.

        \paragraph{Chains of bridges.}
        Now consider chains of bridges. We will first give an argument that
        the leading term in powers of $\beta \Lambda$ is entirely
        momentum independent for a one-loop bridge, before generalizing this
        to chains of arbitrary length.
        In each case, we use the results of the previous section to
        drop terms containing more than a single loop.
        
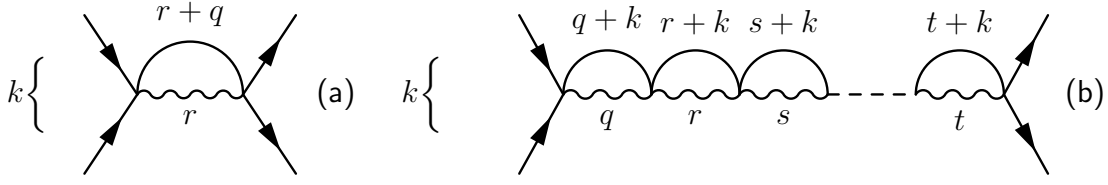
\begin{figure}
                \hfill
                $\displaystyle k \Bigg\{$
                \parbox{33mm}{
                \begin{fmfgraph*}(100,60)
                        \fmfleft{l1,l2}
                        \fmfright{r1,r2}
                        \fmf{fermion}{l1,v1}
                        \fmf{fermion}{l2,v1}
                        \fmf{boson,label=$r$}{v1,v2}
                        \fmf{fermion}{v2,r1}
                        \fmf{fermion}{v2,r2}
                        \fmffreeze
                        \fmf{plain,left,label=$r+q$}{v1,v2}
                \end{fmfgraph*}}
                \figlabel{a}
                \hfill
                $\displaystyle k \Bigg\{$
                \parbox{80mm}{
                \begin{fmfgraph*}(250,60)
                        \fmfleft{l1,l2}
                        \fmfright{r1,r2}
                        \fmf{fermion}{l1,v1}
                        \fmf{fermion}{l2,v1}
                        \fmf{boson,label=$q$}{v1,v2}
                        \fmf{boson,label=$r$}{v2,v3}
                        \fmf{boson,label=$s$}{v3,v4}
                        \fmf{dashes}{v4,v5}
                        \fmf{boson,label=$t$}{v5,v6}
                        \fmf{fermion}{v6,r1}
                        \fmf{fermion}{v6,r2}
                        \fmffreeze
                        \fmf{plain,left,label=$q+k$}{v1,v2}
                        \fmf{plain,left,label=$r+k$}{v2,v3}
                        \fmf{plain,left,label=$s+k$}{v3,v4}
                        \fmf{plain,left,label=$t+k$}{v5,v6}
                \end{fmfgraph*}}
                \figlabel{b}
                \hfill
                \mbox{}
                \caption{\label{fig:single-bridge}Bridge diagrams. In
                \figlabel{a}, $2 \rightarrow 2'$ fermion scattering is dressed
                by a single dark energy bridge, in which a single dark energy
                particle is emitted or absorbed at the scattering vertices.
                In \figlabel{b}, a chain of bridges is exchanged.
                In principle, more complicated configurations exist in which
                the bridge is itself built out of rainbows of sub-bridges.
                We neglect these, since we anticipate that they will be
                suppressed by extra powers of the phase-space factor
                $1/8\pi^2$.}
        \end{figure}

        The single bridge is shown in Fig.~\ref{fig:single-bridge}\figlabel{a}.
        We assume that it describes a bridged fermion scattering process,
        between two fermion species which couple conformally
        to dark energy via the coupling function $F(\beta_f \chi)$
        and coupling $\beta_f$.
        (The generalization to different couplings and
        coupling functions is obvious.)
        It is easy to see that the effect of the bridge is the same as
        inclusion of an extra term in the propagator, corresponding to
        \begin{eqnarray}
                \fl\nonumber
                \bar{F}_1^2 \beta_f^2 \int \frac{\d^4 r}{(2\pi)^4}
                \left( \eta_{ab} + \frac{r_a r_b}{M_W^2} \right)
                \frac{1}{r^2 + M_W^2 - \im \epsilon}
                \frac{1}{(r+q)^2 - \im \epsilon}
                \\ \nonumber
                \simeq
                \frac{\im \bar{F}_1^2 \beta_f^2}{16\pi^2} \eta_{ab}
                \int_0^1 \d x \;
                \Bigg( - \frac{1}{2} \frac{\Lambda^2}{\Lambda^2 + \Delta^2}
                        + \frac{1}{2} \ln \left[ 1 + \frac{\Lambda^2}{\Delta^2} \right]
                        \\ \hspace{3.7cm} \mbox{}
                        + \frac{1}{8 M_W^2} \frac{\Lambda^4 + 2 \Lambda^2 \Delta^2}
                        {\Lambda^2 + \Delta^2} - \frac{\Delta^2}{4 M_W^2}
                        \ln \left[ 1 + \frac{\Lambda^2}{\Delta^2} \right] \Bigg) ,
        \end{eqnarray}
        where $\Delta$ satisfies
        \begin{equation}
                \Delta^2 = x(1-x) q^2 + x M_\chi^2 + (1-x) M_W^2 .
        \end{equation}
        Clearly, the leading term in powers of
        $\beta \Lambda$ is momentum independent.
        
        The case of multiple bridges chained together is shown in
        Fig.~\ref{fig:single-bridge}\figlabel{b}.
        One may wonder why it is necessary to consider such chains,
        since we have already argued that diagrams with a large number of
        loops are suppressed by powers of the phase-space factor
        $8\pi^2$. The reason is that to study corrections to the propagator,
        which potentially shift the location of its pole, one must resum
        enough diagrams to capture shifts in the physical mass of the particle.
        Such shifts are not captured at any finite order in perturbation
        theory.
        
        Exchange of the chained bridge with $n$ \emph{internal} vertices
        is equivalent to introducing an
        extra term in the propagator, which takes the form
        \begin{eqnarray}
                \fl\nonumber
                \bar{F}_1^2 \beta_f^2 ( \bar{B}_2 \beta^2 )^n (\im)^n
                \int \frac{\d^4 q}{(2\pi)^4}
                \cdots
                \int \frac{\d^4 t}{(2\pi)^4}
                \frac{1}{q^2 + M_W^2 - \im \epsilon}
                \frac{1}{(q + k)^2 + M_\chi^2 - \im \epsilon}
                \cdots
                \\ \nonumber \mbox{} \times
                \left( \eta_{ab} + \frac{q_a q_b}{M_W^2} \right)
                \left( \eta^{bc}[ q \cdot r - \gamma_2 M_W^2] - q^c r^b \right)
                \\ \nonumber \mbox{} \times
                \left( \eta_{cd} + \frac{r_c r_d}{M_W^2} \right)
                \left( \eta^{dc}[ r \cdot s - \gamma_2 M_W^2] - r^c s^d \right)
                \\ \nonumber \mbox{} \times
                \left( \eta_{ef} + \frac{s_e s_f}{M_W^2} \right) \cdots
                \\ \nonumber \mbox{} \times
                \left( \eta_{gg} + \frac{t_g t_h}{M_W^2} \right) .
        \end{eqnarray}
        In the first contraction, the cubic term $\sim q^3$ cancels out.
        In the second contraction, the cubic term $\sim r^3$ cancels out.
        The same sequence recurs throughout the chain.
        This implies that the only way to bring each momentum integral to
        order $\sim \Lambda^2$ is consider only the \emph{quadratic} term
        from each propagator, which multiplies the momentum-independent
        quantity $\gamma_W M_W^2$.
        Accordingly, one can conclude that the resummed contribution
        at leading order in $\beta \Lambda$ behaves like an extra term
        in the propagator of the form
        \begin{equation}
                \frac{\im}{64 \pi^2} \bar{F}_1^2 (\beta_f \Lambda)^2 \eta_{ab}
                \frac{1}{M_W^2}
                \frac{1}{1-
                        \frac{\bar{B}_2 \gamma_2 [ \beta \Lambda]^2}{64\pi^2}
                } .
        \end{equation}
        Once again, this term is momentum-independent
        and contributes only to an unobservable shift---the
        same for all species---in the relevant
        masses and coupling constants.

        \section*{References}

			\providecommand{\href}[2]{#2}\begingroup\raggedright\endgroup

\end{fmffile}
\end{document}